\documentclass[preprintnumbers, floatfix, letterpaper, twocolumn,aps,prd,epsfig,nofootinbib,natbib,longbibliography]{revtex4-1}

\usepackage{graphicx}
\usepackage{epstopdf}
\usepackage{latexsym}
\usepackage{amssymb}
\usepackage{amsmath}
\usepackage{color}
\usepackage{mathrsfs}
\usepackage{xparse}
\usepackage{bbding}
\usepackage{pifont}
\usepackage{comment}
\usepackage{ulem}
\usepackage{float}
\usepackage[inline]{enumitem}
\usepackage{soul}
\usepackage{scalerel}
\usepackage[caption=false]{subfig}

\let\Right\right
\let\Left\left
\makeatletter
\def\right#1{\Right#1\@ifnextchar){\!\right}{}}
\def\left#1{\Left#1\@ifnextchar({\!\left}{}}
\makeatother

\usepackage[
            pdfstartview=FitH,
            bookmarksnumbered=true,
            bookmarksopen=true,
            colorlinks,
            linkcolor=blue,
            anchorcolor=green,
            citecolor=blue
            ]{hyperref}
\begin{document}

  \renewcommand\arraystretch{2}
 \newcommand{\bq}{\begin{equation}}
 \newcommand{\eq}{\end{equation}}
 \newcommand{\bqn}{\begin{eqnarray}}
 \newcommand{\eqn}{\end{eqnarray}}
 \newcommand{\nb}{\nonumber}
 \newcommand{\lb}{\label}
 
\newcommand{\La}{\Lambda}
\newcommand{\va}{\scriptscriptstyle}
\newcommand{\be}{\nopagebreak[3]\begin{equation}}
\newcommand{\ee}{\end{equation}}

\newcommand{\ba}{\nopagebreak[3]\begin{eqnarray}}
\newcommand{\ea}{\end{eqnarray}}

\newcommand{\la}{\label}
\newcommand{\n}{\nonumber}
\newcommand{\su}{\mathfrak{su}}
\newcommand{\SU}{\mathrm{SU}}
\newcommand{\U}{\mathrm{U}}
\newcommand{\red}{ }

\newcommand{\R}{\mathbb{R}}

 \newcommand{\cb}{\color{blue}}
    \newcommand{\cc}{\color{cyan}}
        \newcommand{\cm}{\color{magenta}}
\newcommand{\rc}{\rho^{\scriptscriptstyle{\mathrm{I}}}_c}
\newcommand{\rd}{\rho^{\scriptscriptstyle{\mathrm{II}}}_c} 
\NewDocumentCommand{\evalat}{sO{\big}mm}{%
  \IfBooleanTF{#1}
   {\mleft. #3 \mright|_{#4}}
   {#3#2|_{#4}}%
}
\newcommand{\PRL}{Phys. Rev. Lett.}
\newcommand{\PL}{Phys. Lett.}
\newcommand{\PR}{Phys. Rev.}
\newcommand{\CQG}{Class. Quantum Grav.}


\title{Dirac observables in the 4-dimensional phase space  of Ashtekar's variables and spherically symmetric  loop quantum black holes}

\author{Geeth  Ongole  ${}^{a}$}
\email{geeth$\_$ongole1@baylor.edu}

\author{Hongchao Zhang ${}^{b, c}$}
\email{zhanghongchao852@live.com}

\author{Tao Zhu${}^{b, c}$}
\email{zhut05@zjut.edu.cn}

\author{Anzhong Wang${}^{a}$ \footnote{Corresponding author}}
\email{anzhong$\_$wang@baylor.edu; Corresponding author}

\author{Bin Wang${}^{d,e}$}
\email{wang$\_$b@sjtu.edu.cn}

\affiliation{${}^{a}$ GCAP-CASPER, Physics Department, Baylor University, Waco, Texas 76798-7316, USA\\
${}^{b}$Institute for Theoretical Physics \& Cosmology, Zhejiang University of Technology, Hangzhou, 310023, China\\
${}^{c}$ United Center for Gravitational Wave Physics (UCGWP),  Zhejiang University of Technology, Hangzhou, 310023, China\\
 ${}^{d}$ Center for Gravitation and Cosmology, College of Physical Science and Technology, Yangzhou University, Yangzhou, 225009, China\\
 ${}^{e}$ Shanghai Frontier Science Center for Gravitational Wave Detection, Shanghai Jiao Tong University, Shanghai 200240, China}

\date{\today}

\begin{abstract}

In this paper, we study a proposal put forward recently by Bodendorfer, Mele and M\"unch and Garc\'\i{}a-Quismondo and Marug\'an, in which the two 
polymerization parameters of spherically symmetric black hole spacetimes are the Dirac observables of the four-dimensional Ashtekar's variables. In this 
model, black and white hole  horizons in general exist and naturally divide the spacetime into  the external and internal regions. In the external region, 
the spacetime can be made asymptotically flat by properly choosing the dependence of the two polymerization parameters on the Ashtekar variables. Then, 
we find that the asymptotical behavior of the spacetime is universal, and, to the leading order, the curvature invariants are independent of the mass parameter $m$. For 
example, the Kretschmann scalar approaches zero as $K \simeq A_0r^{-4}$ asymptotically, where $A_0$ is generally a non-zero constant and independent 
of $m$, and $r$ the geometric radius of the two-spheres. In the internal region, all the physical quantities are finite, and the Schwarzschild black hole singularity 
is replaced by a transition surface whose radius is always finite and non-zero. The quantum gravitational effects are negligible near the black hole horizon for 
very massive black holes. However, the behavior of the spacetime across the transition surface is significantly different from all loop quantum black holes 
studied so far. In particular, the location of the maximum amplitude of the curvature scalars is displaced from the transition surface and depends on $m$, 
so does the maximum amplitude.  In addition, the  radius of  the white hole is much smaller than that of the black hole, and its exact value sensitively depends 
on $m$, too.

\end{abstract}

\maketitle

\section{Introduction} 
\label{sec:Intro}
 \renewcommand{\theequation}{1.\arabic{equation}}\setcounter{equation}{0}

Loop quantum gravity (LQG)   has burgeoned in an effort to quantize gravity. It is a non-perturbative and background independent approach to canonically quantizing Einstein's general relativity (GR) 
\cite{AALQGreport,Thiemann:2007zz,GP11,MB11,RV15}.
Loop quantum cosmology (LQC) is an application of the LQG techniques by first performing the symmetry reduction of the homogeneous and isotropic spacetimes at the classical level, and then 
quantizing it by using the canonical Dirac quantization for systems with constraints, the so-called minisuperspace approach  \cite{AS11}.
Singularities are one of the major predictions by GR, which appear (classically)  in the very early cosmological epoch  and the interior regions of black holes. Classical GR becomes invalid when such singularities appear.
One usually expects that in such high curvature regimes quantum gravitational effects will take over and become dominant, whereby the singularities are smoothed out and finally replaced by regions with the Planck scale
curvatures. Because of the quantum nature of geometry in LQG, cosmological singularities can be naturally resolved in LQC models, without any additional constraints on matter fields  \cite{AS11}.
Although the full theory is still under construction,   symmetry reduced models constructed from LQG have received great attention.

Since the Schwarzschild interior is isometric to the homogeneous but anisotropic (vacuum) Kantowski-Sachs cosmological model, techniques of LQC can be used to study black hole(BH) singularities in the spherically symmetric spacetimes. In the treatment of LQC, the full quantum evolution is well approximated by  quantum corrected effective equations. Similar treatment is applied to the interior of the Schwarzschild spacetime to get the quantum corrected Schwarzschild spacetime,
 which cures the black hole singularity.  {Recently, such works have received lot of attention}  \cite{Ashtekar:2005qt,Modesto:2005zm,BV07,GP08,Campiglia:2007pb,Brannlund:2008iw,Modesto:2008im,Chiou:2008nm,Chiou:2008eg,Perez:2012wv,Gambini:2013exa,GOP14,GP14,Haggard:2014rza,Joe:2014tca,Corichi:2015xia,Dadhich:2015ora,Olmedo:2017lvt,Cortez:2017alh,CR17,AP17,BMM18,RMD18,BCDHR18,BMM19a,MDR19,Assanioussi:2019twp,AAN20,Ashtekar20,Zhang:2020qxw,Gambini:2020nsf,GSSW2020,Kelly:2020uwj,Liu:2020ola,BMM21a,Giesel:2021dug,BMM21,SG21,GOP21,Li:2021snn,LFSW21,Giesel:2021rky,HL22,ZMSZ22,RD22,Gan22}.
 
A particular model proposed recently  is the Ashtekar-Olmedo-Singh (AOS) loop quantum black hole (LQBH) \cite{AOS18a,AOS18b,AO20}, in which  AOS constructed the effective Hamiltonian that governs the dynamics of 
spherically symmetric loop quantum black holes in the semi-classical limit. This effective Hamiltonian contains  two  polymerization parameters $\left(\delta_b, \delta_c\right)$, characterizing the quantum gravitational effects.   In some of the previous approaches,
they were simply taken as constants\cite{Ashtekar:2005qt,Assanioussi:2019twp,Campiglia:2007pb, BMM19a}, similar to the  $\mu_0$ scheme first introduced in LQC \cite{AS11}. However, in LQC  it was
found  \cite{APS06} that the $\mu_0$ scheme leads to large quantum geometric effects even in  {regions} much lower than the Planck curvatures. To remand this problem,
Ashtekar, Pawlowski and Singh (APS) \cite{APS06} proposed that  the    polymerization parameter should depend on  phase variables, the so-called $\bar\mu$ scheme\footnote{It should be noted that
in LQC, there exists only one parameter $\bar\mu$ corresponding to the area operator  $p$, and APS set it to  $\bar\mu^2 |p| =  \left(4\sqrt{3} \pi\gamma\right)\ell_p^2
\equiv \Delta$ \cite{APS06}, where $\ell_p$ denotes the Planck scale.}.  It turns out that so far this is the only scheme that leads to consistent 
results in LQC \cite{AS11}. 

 {On the other hand,  in the AOS model} \cite{AOS18a,AOS18b,AO20}, instead of treating $\left(\delta_b, \delta_c\right)$ as arbitrary functions of the phase variables,  {they consider them} as Dirac observables, that is, they are particular functions of  the phase variables, such that along the trajectories of the effective Hamiltonian equations they  {become} constants. Similar treatments have also been adopted in  \cite{BV07,Chiou:2008nm,Chiou:2008eg,Corichi:2015xia, Olmedo:2017lvt}.  {But the AOS approach is different as they considered} $\left(\delta_b, \delta_c\right)$ as Dirac observables in the 8-dimensional extended phase
space $\Gamma_{\text{ext}}$ of the variables $\left(b, c, p_b, p_c; \delta_b, \delta_c, p_{\delta_b}, p_{\delta_c}\right)$, instead of the  4-dimensional  phase
space $\Gamma$ of the variables $\left(b, c, p_b, p_c\right)$.   {Another key feature that differentiates the AOS approach is the imposition of the minimum area condition of LQG  on the plaquettes that tesselate the 
transition surface.} This treatment helped resolve the long standing problems in LQBH such as the dependence of the system on the fiducial structure and non-negligible quantum corrections at low curvatures, to name a few.

Despite  {the success of the AOS model, some questions have been raised} \cite{BBCCY20,MB20}. In particular,   Bodendorfer, Mele and M\"unch (BMM) \cite{BMM19} argued that the polymerization parameters 
can be treated canonically as  Dirac observables directly in the 4-dimensional  phase space $\Gamma$, so that $\delta_i = f_i\left(O_i\right), \; (i = b, c)$, where $O_i$'s are the two independent Dirac observables that can be constructed in the spherically symmetric spacetimes, and are given explicitly by Eq.(\ref{Ob}) below in terms of the four Ashtekar variables $(b, c, p_b, p_c)$. Then, the corresponding dynamics of the effective Hamiltonian is different from that of AOS. 
More recently,  Garc\'\i{}a-Quismondo and Marug\'an (GM) \cite{GQ21} argued that in the BMM approach, the two polymerization parameters in general should depend on both $O_b$ and $O_c$, that is, $\delta_i = f_i\left(O_b, O_c\right)$, and 
the BMM choice  {can be realized as a special case. GM also derived the corresponding dynamical equations.}

In this paper, we shall study the main properties of the LQBH spacetimes resulting from the BMM/GM proposal. In particular, the paper will be organized as follows: In Section \hyperref[Sec:II]{II} we will briefly review the AOS model, so readers can clearly see the difference 
between the AOS and BMM/GM approaches. In Section \hyperref[Sec:III]{III}, we first introduce the  BMM/GM model and   then restrict ourselves to the external region of the BMM/GM LQBH spacetime. By requiring that the spacetime in this region be asymptotically flat, we find that
the parameter $\Omega_b \; [\equiv \omega_{bb} + \omega_{bc}]$ must be non-negative $\Omega_b \ge 0$,  where $\omega_{ij} \equiv \partial f_i/\partial O_j$  [cf. Eq.(\ref{eq3.13})]. This excludes the BMM choice $\delta_i = f_i\left(O_i\right)$ \cite{BMM19}, which is also  the choice made by AOS \cite{AOS18a,AOS18b,AO20}, but it must be noted that AOS did it in the extended phase space. With this condition, we find that the asymptotical behavior of the spacetime is universal and independent of the 
mass  parameter $m$ for the curvature invariants [cf. Eqs.(\ref{eq3.24}) and (\ref{eq3.29})]. In particular,  the Kretschmann scalar behaves as   $ K \rightarrow A_0r^{-4}$ as $r \rightarrow \infty$, where $A_0$ is a constant and independent of $m$, and $r$ the geometric radius of the two-spheres. Similar behavior is also found in the AOS model. 

In Section \hyperref[Sec:IV]{IV}, we analyze the properties of the BMM/GM model in the internal region and find that all the physical quantities are finite, and the Schwarzschild black hole singularity is replaced by a transition surface whose radius is always finite and non-zero. 
However, the behavior of the spacetime across the transition surface is significantly different from all LQBHs studied previously. In particular,  
the curvature invariants, such as the  Kretschmann scalar, achieve their maxima  not at the transition surface but right after or before crossing it.  Detailed investigations of the metric components   reveal that this is  because of the  fact that 
now  $\delta_i$'s  are the Dirac observables in  the 4D phase space, which
considerably modify   the structure of the spacetime.  Due to such modifications, the location of the white hole horizon is also very near to the transition surface, and the ratio of the white and black hole horizon radii is much smaller than one,
 and sensitively depends on 
the mass parameter $m$.   Finally, in Section \hyperref[Sec:V]{V}, we summarize our main conclusions.  

To distinguish the AOS and BMM/GM approaches, in this paper, we shall refer them as to the {\it extended and canonical phase space approaches}, respectively. 

Before proceeding further, we would also like to note that parts of the results presented in this paper had
been reported in the APS April meeting, April 9 - 12, 2022, New York, as well as in the 23rd International Conference on General Relativity and Gravitation (GR23),   Liyang, China, July 3 - 8, 2022.

\section{Extended Phase Space Approach} 
\label{Sec:II}
 \renewcommand{\theequation}{2.\arabic{equation}}\setcounter{equation}{0}

The starting point of LQG is the introduction of the Ashtekar variables. In the spherically symmetric spacetimes, they are the metric components $p_b$ and $p_c$ and their moment conjugates $b$ and $c$
with the  canonical relations
\bq
\lb{CRs}
\{b,p_b\}=G\gamma, \quad \{c,p_c\}=2G\gamma,
\eq
where  $\gamma$ is the Barbero-Immirzi parameter and $G$ is the Newtonian gravitational constant.  

In terms of $p_b$ and $p_c$, the four-dimensional spacetime {line element} takes the form, 
\bq
\lb{AOS2e}
ds^2 = - N^2 dT^2 + \frac{p_b^2}{|p_c| L_o^2} dx^2 + |p_c|d\Omega^2,
\eq
where $N$ is the lapse function, and  $L_o$ is a constant, denoting the length of the fiducial cell in the $x$-direction with $x \in \left(0, L_o\right)$, and 
 {$d\Omega^2 \equiv d\theta^2 + \sin^2\theta d\phi^2$} with $\theta$ and $\phi$ being the two angular coordinates defined on the two spheres $T, x = $ Constant. 

In the internal region of a classical black hole, $(N, p_b, p_c)$ are all functions of $T$ only (so are $b$ and $c$), and the corresponding spacetimes are of the Kantowski-Sachs cosmological model, which allows one to apply  {LQC techniques} to such homogeneous but anisotropic spacetimes. As a result, the internal region of the Schwarzschild has been extensively studied in the framework of LQC.  

On the other hand, 
in the external region, the coordinates $T$ and $x$ exchange their rules, 
and the spacetime becomes static. However, such changes can be also carried out by the replacement $N \rightarrow i N$ and $p_b  \rightarrow i p_b$, as shown explicitly below, while 
keeping the dependence of the Ashtekar variables still only on $T$. 

With the above in mind, we can see that in general the metric (\ref{AOS2e}) has the gauge freedom,
\bq
\lb{FD}
T' = T'(T), \quad x' = \alpha x + x_0,  
\eq
in both external and internal regions, where $T'(T)$ is an arbitrary function of $T$ only, and $\alpha$ and $t_0$ are real constants. 
To see the AOS approach  more clearly, let us consider the AOS effective Hamiltonian inside and outside the LQBH,  separately.

\subsection{  AOS Internal Solution}

With the gauge freedom of (\ref{FD}), AOS chose $T'(T)$ so that 
\bq
\label{lapse}
N= \frac{\gamma \delta_b\; {\text{sgn}}\left(p_c\right) \sqrt{|p_c|}}{\sin{\left(\delta_b b\right)}}.
\eq
Then, the effective Hamiltonian in the   interior of LQBHs reads  \cite{AOS18a,AOS18b,AO20} 
\begin{align}
\begin{aligned}
\label{Heff}
H_{\mathrm{eff}}  = -\frac{1}{2 G \gamma} \Bigg[ & 2 \frac{\sin \left(\delta_{c} c\right)}{\delta_{c}} \left|p_{c}\right|  \\
 & + \left(\frac{\sin \left(\delta_{b} b\right)}{\delta_{b}}+\frac{\gamma^{2} \delta_{b}}{\sin \left(\delta_{b} b\right)}\right) p_{b} \Bigg],
\end{aligned}
\end{align}
where    $\delta_b$ and $\delta_c$ are two Dirac observables, appearing in the polymerizations
\bq
\lb{PZ}
b \rightarrow \frac{\sin\left(\delta_b b\right)}{\delta_b}, \quad c \rightarrow \frac{\sin\left(\delta_c c\right)}{\delta_b}.
\eq
That is, replacing $b$ and $c$ by Eq.(\ref{PZ}) in the classical Hamiltonian
\bq
\lb{CH}
H_{\text{cl}} = -\frac{1}{2 G \gamma} \Bigg[ 2c\left|p_{c}\right|    + \left(b+\frac{\gamma^{2}}{b}\right) p_{b}\Bigg],
\eq
whereby  the effective Hamiltonian (\ref{Heff}) is obtained, provided that the classical lapse function is chosen as 
\bq
\lb{CL}
N_{\text{cl}} = \frac{\gamma\; {\text{sgn}}\left(p_c\right) \sqrt{|p_c|}}{b}.
\eq

To fix $\delta_b$ and $\delta_c$, AOS first noticed that the above effective Hamiltonian can be written as
\begin{align}
\label{PHeff}
H_{\mathrm{eff}} &=\frac{L_{o}}{G}\left(O_{b}-O_{c}\right), 
\end{align}
where
\bqn
\label{Ob}
    O_{b} &\equiv&-\frac{p_{b}}{2 \gamma L_{o}}\left(\frac{\sin \left( \delta_{b} b\right)}{\delta_{b}}+\frac{\gamma^{2} \delta_{b}}{\sin \left(\delta_{b} b\right)}\right), \\
\label{Oc}
    O_{c} &\equiv&\frac{\left|p_{c}\right|}{\gamma L_{o}} \frac{\sin\left( \delta_{c} c\right)}{\delta_{c}},
\eqn
are two Dirac observables.  Then, AOS proceeded as follows:  

\begin{itemize}

\item First extend the 4-dimensional (4D) phase space $\Gamma$ spanned by $\left(b, c; p_b, p_c\right)$ to  8-dimensional (8D) phase space $\Gamma_{\text{ext}}$
spanned by $\left(b, c, \delta_b, \delta_c; p_b, p_c, p_{\delta_b},  p_{\delta_c}\right)$. In $\Gamma_{\text{ext}}$ the variables  $\delta_b$ and $\delta_c$ are independent, so they are in particular not functions of $\left(b, c, p_b, p_c\right)$
and instead Poisson commute with all of them. 

\item Lift $H_{\mathrm{eff}}$ given by Eq.(\ref{Heff}) to $\Gamma_{\text{ext}}$, and then consider its Hamiltonian flow. Since $O_{b}$ and $O_{c}$ are the Dirac observables of this flow, 
 {the following choice can be made}
\bq
\lb{Obc}
\delta_b = \delta_b\left(O_b\right), \quad \delta_c = \delta_c\left(O_c\right),  
\eq
so that $\left(\delta_b, \delta_c\right)$ are also the Dirac observables. 

\item Introduce these dependences as two new first-class constraints
\bqn
\lb{Obc1}
&& \Phi_b \equiv O_b  - F_b\left(\delta_b\right)  \simeq 0, \nb\\
&& \Phi_c \equiv O_c  - F_c\left(\delta_c\right)  \simeq 0,   
\eqn
so that the four-dimensional reduced $\hat\Gamma$ corresponding to these constraints is symplectomorphic to the original phase space $\Gamma$. Since $O_b$ and $O_c$ are the Dirac observables, Eq.(\ref{Obc1})  {implies}
\bq
\lb{Obc2}
\delta_b = F_b^{-1}\left(O_b\right), \quad \delta_c =  F_c^{-1}\left(O_c\right),
\eq
 are also constants on the trajectories of the effective Hamiltonian $H_{\mathrm{eff}}$ given by Eq.(\ref{Heff}). 
 
\item To fix  $\delta_b$ and $\delta_c$,  AOS assumed that at the  {transition surface}, (where $T = {\cal{T}}$), the physical areas of the 
($x, \theta$)- and ($\theta, \phi$)-planes are respectively equal to the minimal area $\Delta$ \cite{AOS18a}
\bqn
\lb{eqAOS3a}
 && 2\pi \delta_c \delta_b \left|p_b({\cal{T}})\right|  = \Delta, \\
 \lb{eqAOS3b}
 && 4\pi  \delta_b^2 p_c({\cal{T}})  = \Delta. 
\eqn

\end{itemize}

With all the above, AOS found that the corresponding Hamilton equations are given by
\bqn
\lb{eq2.12}
\dot{b} &=& - \frac{1}{2}\left(\frac{\sin\left(\delta_b b\right)}{\delta_b} + \frac{\gamma^2\delta_b}{\sin\left(\delta_b b\right)}\right), \\
\lb{eq2.13}
\dot{p}_b &=& \frac{1}{2}p_b\cos\left(\delta_b b\right)\left(1  -  \frac{\gamma^2\delta_b^2}{\sin^2\left(\delta_b b\right)}\right),
\eqn
and 
\bqn
\lb{eq2.14}
\dot{c} &=& - 2 \frac{\sin\left(\delta_c c\right)}{\delta_c}, \\
\lb{eq2.15}
\dot{p}_c &=&{2}p_c\cos\left(\delta_c c\right).
\eqn
It is remarkable to note that in the above equations, no cross terms exists between the equations for $(b, p_b)$ and the ones for 
 $(c, p_c)$. As a result, we can solve the two sets of equations independently, and the corresponding solutions are given by  \cite{AOS18a,Gan22} 
\bqn
\lb{AOS2ba}
  \cos\left(\delta_b b\right) 
&=&  b_o\frac{1+ b_{o} \tanh\left(\frac{b_{o} T}{2}\right)}{b_o + \tanh\left(\frac{b_{o} T}{2}\right)}, \nb\\
&=& b_o \frac{b_+ e^{b_oT} - b_-}{b_+ e^{b_oT} + b_-}, \nb\\
p_b &=& - \frac{mL_o}{2b_o^2}\left(b_+  + b_- e^{-b_oT}\right){\cal{A}},\\
\lb{AOS2bb}
 \sin\left(\delta_c c\right) &=& \frac{2a_oe^{2T}}{a^2_o + e^{4T}}, \nb\\
 p_c &=& 4m^2\left(a_o^2 + e^{4T}\right)e^{-2T}, 
\eqn
where  
\bqn
\lb{AOS2c}
{\cal{A}} &\equiv& \Big[2\left(b_o^2 + 1\right)e^{b_oT} - b_-^2 - b_+^2 e^{2b_oT}\Big]^{1/2}, \nb\\  
a_o &\equiv&  \frac{\gamma \delta_c L_o}{8m}, \quad b_o \equiv \left(1 + \gamma^2 \delta_b^2\right)^{1/2}, \nb\\
b_{\pm} &\equiv& b_o \pm 1,
\eqn
with
\bqn
\lb{AOS2d}
&& \delta_b b \in \left(0, \pi\right), \quad \delta_c c \in \left(0, \pi\right),\nb\\
&& p_b \le 0, \quad p_c \ge 0, \quad - \infty < T < 0.
\eqn
The parameter $m$ is an integration constant, related to the mass parameter of the AOS solution. From the above solution, it can be shown that the two Dirac observables on-shell  are
given by
\bqn
\lb{eq2.25}
O_b = m = O_c .  
\eqn

In   the large mass limit, $m \gg m_{p}$, from Eqs.(\ref{eqAOS3a}) and (\ref{eqAOS3b}) AOS found that 
\bqn
\lb{eqAOS34}
 \delta_b = \left(\frac{\sqrt{\Delta}}{\sqrt{2\pi} \gamma^2 m}\right)^{\frac{1}{3}}, \;\;\;
  L_0 \delta_c = \frac{1}{2} \left(\frac{\gamma \Delta^2}{4\pi^2 m}\right)^{\frac{1}{3}}, ~~~~~
\eqn
where $m_p$ denotes the Planck mass.

It should be noted that in \cite{AOS18a} two solutions for $c$ were given, and here in this paper we only consider the one with ``+" sign, as physically they describe the same spacetime.

\begin{figure}[h!]
\includegraphics[height=8.5cm]{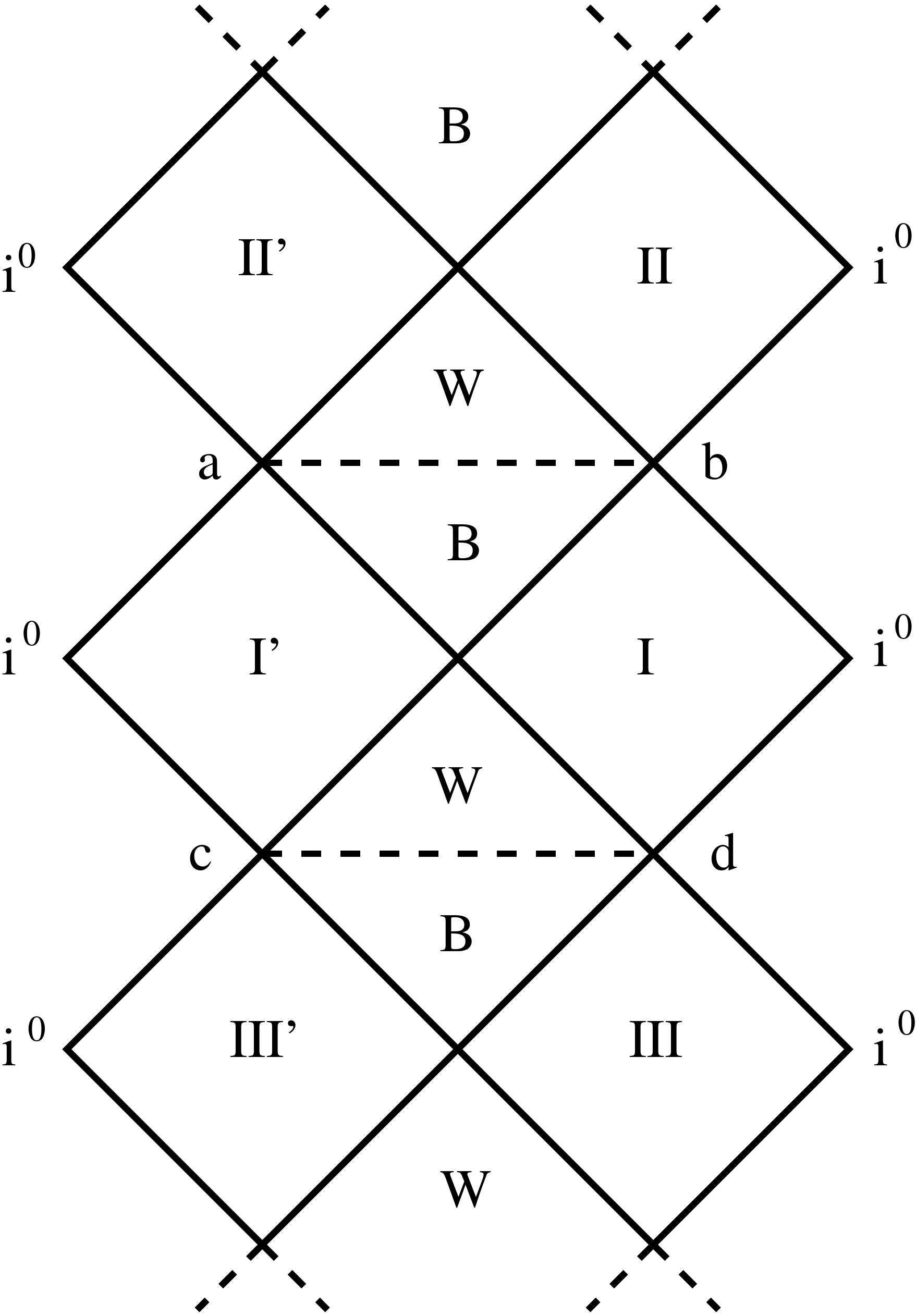}
\caption{The Penrose diagram for the AOS LQBH. The dashed horizontal lines $ab$ and $cd$ represent the transition surfaces (throats), and the regions marked with B  {is the BH interior, and 
the regions marked with W is the WH interior}, but there are no spacetime singularities, so the extensions are infinite along the vertical line in both directions.  Regions marked with I, I', II, II', III, and III'
are asymptotically flat regions but with a falling rate slower than that of the Schwarzschild black hole \cite{AO20}.} 
\label{fig1}
\end{figure}

From Eq.(\ref{AOS2bb}), it can be seen that the transition surface is located at $\left. \partial p_c\left(T\right)/\partial T \right|_{T ={\cal{T}} } = 0$, which yields
\bq
\lb{AOS2f}
{\cal{T}} = \frac{1}{2}\ln\left(\frac{\gamma \delta_c L_o}{8m}\right) < 0.
\eq
There also exist two horizons, located respectively at
\bq
\lb{AOS2g}
T_{\text{BH}}  = 0, \quad T_{\text{WH}} = - \frac{2}{b_o}\ln\left(\frac{b_o + 1}{b_o -1}\right), 
\eq
at which we have ${\cal{A}}(T) = 0$, where $ T = T_{\text{BH}}$ is the location of the black hole horizon, while $T = T_{\text{WH}}$ is the location of the white hole. 
In the region ${\cal{T}} < T < 0$, the 2-spheres are all trapped, while in the one  $T_{\text{WH}} < T < {\cal{T}}$, they are all anti-trapped.  Therefore, 
the region ${\cal{T}} < T < 0$ behaves like the  {BH interior}, while the one $T_{\text{WH}} < T < {\cal{T}}$ behaves like the WH interior,
denoted, respectively, by Region B and Region W in Fig. \ref{fig1}. This explains the reason why we call them the black hole and white hole regions, although 
the geometric radius $\sqrt{p_c}$ of the two-sphere ($T, x$ =  Const) is always finite and non-zero, so spacetime singularities never appear. 

Finally, we note that in this region the lapse function reads 
\bqn
\lb{eq2.29}
N &=& \frac{\gamma \delta_b\; \text{sgn}\left(p_c\right)\left|p_c\right|^{1/2}}{\sin\left(\delta_b b\right)}\nb\\
 &=& \frac{2m}{{\cal{A}}}\left(b_+e^{b_oT} + b_-\right) \left(a_o^2e^{-2T} + e^{2T}\right)^{1/2}, ~~~~~~
\eqn
where ${\cal{A}}$ is given in Eq.(\ref{AOS2c}).

\subsection{AOS External  Solution}

 At the two horizons (\ref{AOS2g}), we have ${\cal{A}}(T) = 0$, and the metric becomes singular, so  extensions beyond these surfaces are needed in order to obtain 
 a geodesically complete spacetime.   AOS showed that such extensions can be obtained from  (\ref{Heff})  
 by the following replacements
\begin{align}
\lb{ext}
b \rightarrow i b, \quad p_{b} \rightarrow i p_{b} , \quad c \rightarrow c, \quad p_{c} \rightarrow p_{c},
\end{align}
for which the canonical relations (\ref{CRs}) now become
\bq
\lb{CRse}
\{b,p_b\}= - G\gamma, \quad \{c,p_c\}=2G\gamma,
\eq
while  the effective Hamiltonian in the external space of the LQBH is given by
\bqn
\label{eHeff}
H_{\mathrm{eff}}^{\mathrm{ext}} &=& -\frac{1}{2 G \gamma}\Bigg[2 \frac{\sin \left(\delta_{c} c\right)}{\delta_{c}} p_{c} \nb\\
 & & - \left(\frac{\sinh \left(\delta_{b} b\right)}{\delta_{b}}- \frac{\gamma^{2} \delta_{b}}{\sinh \left(\delta_{b} b\right)}\right) p_{b}\Bigg]\nb\\
 &=& \frac{L_o}{G}\left(O_b - O_c\right),
\eqn
but now with 
\bqn  
\label{Obe}
    O_{b} &\equiv&\frac{p_{b}}{2 \gamma L_{o}}\left(\frac{\sinh\left(\delta_{b} b\right)}{\delta_{b}}-\frac{\gamma^{2} \delta_{b}}{\sinh\left(\delta_{b} b\right)}\right), \\
\label{Oce}
    O_{c} &\equiv&\frac{p_{c}}{\gamma L_{o}} \frac{\sin\left( \delta_{c} c\right)}{\delta_{c}},
\eqn
which can be obtained directly from Eqs.(\ref{Ob}) and (\ref{Oc}) with the replacement (\ref{ext}). Then, the corresponding Hamilton equations
for $(c, p_c)$ are still given by Eqs.(\ref{eq2.14}) and (\ref{eq2.15}), 
while the ones for $(b, p_b)$ now are replaced by
\bqn
\lb{eq2.12a}
\dot{b} &=& - \frac{1}{2}\left(\frac{\sinh\left(\delta_b b\right)}{\delta_b} - \frac{\gamma^2\delta_b}{\sinh\left(\delta_b b\right)}\right), \\
\lb{eq2.13a}
\dot{p}_b &=& \frac{1}{2}p_b\cosh\left(\delta_b b\right)\left(1  +  \frac{\gamma^2\delta_b^2}{\sinh^2\left(\delta_b b\right)}\right).
\eqn

Then, the corresponding solutions of the Hamilton equations are given by 
\bqn
\label{eq:eb}
 \cosh \left(\delta_{b} b\right)&=& b_o\frac{1+ b_{o} \tanh\left(\frac{b_{o} T}{2}\right)}{b_o + \tanh\left(\frac{b_{o} T}{2}\right)},\nb\\
 p_{b} &=&  -2m\gamma L_o \delta_b  \frac{\sinh\left(\delta_{b} b \right)}{\gamma^2\delta_{b}^2 - \sinh^2\left(\delta_{b} b \right)}\nb\\
&=&  - \frac{mL_o}{2b_o^2}\left(b_+ + b_-e^{-b_o T}\right){\cal{A}},\\
\label{ec}
\sin\left(\delta_c c\right) &=& \frac{2a_oe^{2T}}{a^2_o + e^{4T}}, \nb\\
   p_{c}&=& 4 m^{2}\left(e^{2 T}+a_{o}^2 e^{-2 T}\right),
\eqn
but now with
\bq
\lb{eq2.26}
{\cal{A}} \equiv \Big[b_-^2 + b_+^2 e^{2b_oT}-2\left(b_o^2 + 1\right)e^{b_oT}\Big]^{1/2},
\eq
which can be obtained from Eq.(\ref{AOS2c}) by the replacement ${\cal{A}} \rightarrow  i {\cal{A}}$ (or ${\cal{A}}^2 \rightarrow - {\cal{A}}^2$),
so that $g_{xx} \rightarrow - g_{xx}$, and the coordinate $x$ now becomes timelike in the external region ($T > 0$) of the black hole horizon,
located at $T = 0$.
It can be shown that for the above solution, we have
$O_b  = m =  O_c  $,
which shows clearly that $O_b$ and $O_c$ defined by Eqs.(\ref{Obe}) and (\ref{Oce}) are two Dirac observables.

We also note that the replacement of Eq.(\ref{ext}) leads to
\bqn
\lb{AOS2fa}
N^2 &=& -  \frac{\gamma^2 \delta_b^2\left|p_c\right|}{\sinh^2\left(\delta_b b\right)}\nb\\
&=& - \frac{4m^2}{{\cal{A}}^2}\left(b_+e^{b_oT} + b_-\right)^{2} \left(a_o^2e^{-2T} + e^{2T}\right), ~~~~~
\eqn
so that,  in terms of $N$, $p_b$ and $p_c$, the metric now takes the form \cite{AOS18a}
\bqn
\lb{eq2.41}
ds^2 &=& - N^2 dT^2 - \frac{p_b^2}{|p_c| L_o^2} dx^2 + |p_c|d\Omega^2\nb\\
&=& - \frac{p_b^2}{|p_c| L_o^2} dx^2 + \frac{\gamma^2 \delta_b^2\left|p_c\right|}{\sinh^2\left(\delta_b b\right)} dT^2  + |p_c|d\Omega^2, ~~~~~
\eqn
which shows clearly that now $T$ is spacelike, while $x$ becomes timelike, so the spacetime outside of the LQBH is static.

In addition, AOS showed that the two metrics (\ref{AOS2e}) and (\ref{eq2.41}) are analytically connected  {to each other} across the two horizons, and as a result, 
the extensions are unique. The global structure of the spacetime is given by the Penrose diagram of Fig. \ref{fig1}, from which we can see that the extensions along the vertical direction 
are infinite, quite similar to the charged spherically symmetric Reissner-Nordstr\"om solutions \cite{HE73}, but without spacetime singularities, as now the 
geometric radius $\sqrt{p_c}$ never becomes zero. 

Before proceeding to the next section, we also note that technically the AOS extended space approach can be realized directly by taking $\delta_b$ and $\delta_c$ to be constants in the
phase space of $(b, c, p_b, p_c)$, and then impose the conditions (\ref{eqAOS3a}) and (\ref{eqAOS3b}), as by definition constants over the whole phase space  are also Dirac observables.

\section{Canonical Phase Space Approach} 
\label{Sec:III}
 \renewcommand{\theequation}{3.\arabic{equation}}\setcounter{equation}{0}

Instead of extending the 4D physical phase space to 8D phase space,  and then considering $\delta_b$ and $\delta_c$ as the Dirac observables of the extended phase space,
Bodendorfer, Mele, and M\"unch (BMM) pointed out \cite{BMM19} that they can be  considered directly as the Dirac observables in the 4D physical phase space of $(b, c, p_b, p_c)$,
as those given by Eq.(\ref{Obc}). Lately,  Garc\'\i{}a-Quismondo and Marug\'an argued \cite{GQ21} that $\delta_b$ and $\delta_c$ should in general depend on both of the two Dirac
observables $O_b$ and $O_c$,
\bq
\lb{eq3.1}
\delta_i = f_i\left(O_b, O_c\right), \; (i = b, c),
\eq
while  Eq.(\ref{Obc}) only represents a particular choice of the  general case. Eq.(\ref{eq3.1}) shows clearly that now $\delta_b$ and $\delta_c$ all depend on
the four variables, $(b, c, p_b, p_c)$, through Eqs.(\ref{Ob}) and (\ref{Oc}) [or Eqs.(\ref{Obe}) and (\ref{Oce}) when outside of the LQBH]. Then, the corresponding Hamilton equations are given by \cite{GQ21}
\bqn
\lb{eq3.2a}
\partial_{T} i&=& C_{ij} \left[s_{i} \frac{L_{o}}{G}\left\{i, p_{i}\right\} \frac{\partial O_{i}}{\partial p_{i}}\right],\\
\lb{eq3.2b}
\partial_{T} p_i&=& C_{ij} \left[-s_{i} \frac{L_{o}}{G}\left\{i, p_{i}\right\} \frac{\partial O_{i}}{\partial i}\right],
\eqn
where $i,j=b,c$, $i\neq j$,  $s_b = 1$, $s_c =-1$, and  
\bqn
\lb{eq3.3a}
    C_{ij}&\equiv& \frac{1-\Delta_{j j}-\Delta_{j i}}{\left(1-\Delta_{i i}\right)\left(1-\Delta_{j j}\right)-\Delta_{i j} \Delta_{j i}},\\
    \lb{eq3.3b}
\Delta_{i j}&\equiv& \frac{\partial O_{i}}{\partial \delta_{i}} \frac{\partial f_{i}}{\partial O_{j}}.
\eqn
It is interesting to note that, introducing two new variables, $t_i, (i = b, c)$, via the relations 
\bq
\lb{eq3.4}
    d t_i \equiv C_{ij} d T,\; (i \not= j),
\eq
Eqs.(\ref{eq3.2a}) and (\ref{eq3.2b}) take the forms,
\bqn
\lb{eq3.2aa}
\partial_{t_i} i&=&  s_{i} \frac{L_{o}}{G}\left\{i, p_{i}\right\} \frac{\partial O_{i}}{\partial p_{i}},\\
\lb{eq3.2bb}
\partial_{t_i} p_i&=&  -s_{i} \frac{L_{o}}{G}\left\{i, p_{i}\right\} \frac{\partial O_{i}}{\partial i}, 
\eqn
which will lead to the same Hamilton equations as those given by AOS, if we replace $T$ by $t_b$ in the equations for $b$ and $p_b$, and
$T$ by $t_c$ in the equations for $c$ and $p_c$, as first noted in \cite{GQ21}.   This observation will significantly simplify our following
discussions. 

To proceed further, in the rest of this section, let us consider the above equations only in the external region, while the ones in the 
internal region will be considered in the next section.

\vspace{.5cm}

\subsection{Dynamics of the external LQBH Spacetimes}

In the external region,  the Hamilton equations take the form
\bqn
\lb{eq3.5a}
\frac{db}{dt_b}   &=& - \frac{1}{2}\left(\frac{\sinh\left(\delta_b b\right)}{\delta_b} - \frac{\gamma^2\delta_b}{\sinh\left(\delta_b b\right)}\right), \\
\lb{eq3.5b}
\frac{d p_b}{dt_b} &=& \frac{1}{2}p_b\cosh\left(\delta_b b\right)\left(1  +  \frac{\gamma^2\delta_b^2}{\sinh^2\left(\delta_b b\right)}\right),
\eqn
for $(b, p_b)$,  and
\bqn
\lb{eq3.6a}
\frac{dc}{dt_c}  &=& - 2 \frac{\sin\left(\delta_c c\right)}{\delta_c}, \\
\lb{eq3.6b}
\frac{dp_c}{dt_c} &=&{2}p_c\cos\left(\delta_c c\right),
\eqn
for  $(c, p_c)$. Then, the corresponding solutions for $b$ and $p_b$ will be given by Eqs.(\ref{eq:eb}) and (\ref{eq2.26}) by simply replacing 
$T$ by $t_b$, that is,
\bqn
\lb{eq3.7}
 && \cosh \left(\delta_{b} b\right) = b_o\frac{1+ b_{o} \tanh\left(\frac{b_{o} t_b}{2}\right)}{b_o + \tanh\left(\frac{b_{o} t_b}{2}\right)},\nb\\
&& p_{b} =  - \frac{mL_o}{2b_o^2}\left(b_+ + b_-e^{-b_o t_b}\right){\cal{A}},\nb\\
&&  {\cal{A}} \equiv \Big[b_-^2 + b_+^2 e^{2b_o t_b}-2\left(b_o^2 + 1\right)e^{b_o t_b}\Big]^{1/2}, ~~~~~~~~~~~~
 \eqn
while the solutions for $c$ and $p_c$ will be given by Eqs.(\ref{ec})  with the replacement 
$T$ by $t_c$, i.e.
\bqn
\lb{eq3.8}
&& \sin\left(\delta_c c\right) = \frac{2a_oe^{2t_c}}{a^2_o + e^{4t_c}}, \nb\\
&&   p_{c} = 4 m^{2}\left(e^{2 t_c}+a_{o}^2 e^{-2 t_c}\right).
\eqn

The relation between $t_b$ and $t_c$ is given by Eq.(\ref{eq3.4}), from which we find that
\bq
\lb{eq3.9}
dt_c = \frac{C_{cb}}{C_{bc}} dt_b.
\eq
To study the above relation, let us first note that  Eqs.(\ref{Obe}) and (\ref{Oce}) lead to
\begin{widetext}
\bqn
\lb{eq3.10}
\frac{\partial O_{b}}{\partial \delta_{b}} &=& \frac{p_{b}}{2 \gamma L_{o}\delta_b^2}\left(1+\frac{\gamma^{2} \delta_{b}^{2}}{\sinh ^{2}\left(\delta_{b}b\right)}\right)
 \big[\left(\delta_{b} b\right) \cosh\left( \delta_{b} b\right)-\sinh \left(\delta_{b}b\right)\big], \nb\\
\frac{\partial O_{c}}{\partial \delta_{c}}&=&\frac{p_{c}}{\gamma L_{o} \delta_c^2} \big[\left(\delta_{c} c\right) \cos\left( \delta_{c} c\right)-\sin\left( \delta_{c} c\right)\big]. 
\eqn
Then, we find that
\bqn
\lb{eq3.11}
C_{b c}&=& \frac{1}{\cal{D}}\left(1- \Omega_{c}\frac{\partial O_{c}}{\partial \delta_{c}} \right)=\frac{1}{\cal{D}}\left\{1- \frac{\Omega_c p_c}{ \gamma L_{o} \delta_{c}^{2}}  \Big[\left(\delta_{c} c\right) \cos \left(\delta_{c} c\right)-\sin \left(\delta_{c} c\right)\Big]\right\} ,\nb\\
C_{cb} &=&\frac{1}{\cal{D}}\left(1- \Omega_{b} \frac{\partial O_{b}}{\partial \delta_{b}} \right) =\frac{1}{\cal{D}}\Bigg\{1- \frac{\Omega_b p_b}{2 \gamma L_{0} \delta_{b}^{2}}\left(1+ \frac{\gamma^2\delta_b^2}{\sinh^2\left(\delta_b b\right)}\right)  \Big[\left(\delta_{b} b\right) \cosh \left(\delta_{b} b\right)-\sinh \left(\delta_{b} b\right)\Big]\Bigg\},
\eqn
where
\bqn
\lb{eq3.12}
{\cal{D}} &\equiv& 1 - \omega_{cc} \frac{\partial O_{c}}{\partial \delta_{c}} -  \omega_{bb} \frac{\partial O_{b}}{\partial \delta_{b}} + \left( \omega_{bb}\omega_{cc}-\omega_{bc}\omega_{cb} \right) \frac{\partial O_{b}}{\partial \delta_{b}} \frac{\partial O_{c}}{\partial \delta_{c}}\nb\\
&=&1 - \frac{ \omega_{cc} p_{c} }{\gamma L_{o} \delta_{c}^{2}}\big[\left(\delta_c c\right) \cos \left(\delta_c c\right)-\sin \left(\delta_c c\right)\big] -  \frac{\omega_{bb} p_{b}}{2 \gamma L_{0} \delta_{b}^{2}}\left(1+\frac{\gamma^{2} \delta_{b}^{2}}{\sinh ^{2}\left(\delta_{b} b\right)}\right)  \left[\left(\delta_{b} b\right) \cosh \left(\delta_{b} b\right)-\sinh \left(\delta_{b} b\right)\right] \nb\\
&& ~ +\frac{\omega_{bb}\omega_{cc}-\omega_{bc}\omega_{cb}}{2\gamma^2L_{o}^2\delta_b^2\delta_c^2}p_bp_c\left(1+\frac{\gamma^{2} \delta_{b}^{2}}{\sinh ^{2}\left(\delta_{b} b\right)}\right) \big[\left(\delta_c c\right) \cos \left(\delta_c c\right)-\sin \left(\delta_c c\right)\big] \left[\left(\delta_{b} b\right) \cosh \left(\delta_{b} b\right)-\sinh \left(\delta_{b} b\right)\right], ~~~ \\
\label{eq3.13}
\omega_{ij} &\equiv& \frac{\partial f_i}{\partial O_j}, \quad 
\Omega_c \equiv \omega_{cc}+\omega_{cb} , \quad \Omega_b \equiv  \omega_{bb}+\omega_{bc}. 
\eqn
It should be noted that the numerator of $C_{bc}$ is a function of $t_c$ and the one of $C_{cb}$ is a function of $t_b$, where $t_b$ and $t_c$ are related one to the other through Eq.(\ref{eq3.9}). In particular, for $t_{b}, t_{c} \gg 1$, from Eq.(\ref{eq3.9})  
we find 
\begin{align}
\label{eq3.20}
    \begin{aligned}
     t_c - 2 \beta m^2 a_o^3 + {\cal{O}}\left(e^{-4 t_c}\right) = \left(1+\alpha_2\right) t_b + \frac{\alpha_1 }{b_o} e^{b_o t_b} + \frac{\alpha_3}{2 b_o}   + {\cal{O}}\left(e^{-b_o t_b}\right),\; \left(t_{b}, t_{c} \gg 1\right), 
    \end{aligned}
\end{align}
\end{widetext}
where
\bqn
\label{eq3.20a}
    \alpha_1 &=& \frac{m (b_o+1)^2}{2 \gamma b_o^2  \delta_b^2} \left(b
   _o \cosh^{-1}{b_o} -\gamma\delta_b\right)\Omega_b, \nb\\
   \alpha_2 &=& -\frac{m \gamma ^2 \delta_b  }{\gamma ^2 \delta_b^2 + 1} \Omega_b,
\eqn
$\beta$ and $\alpha_3$ are other constants, and their explicit expressions will not be given here, as they will not affect our following discussions.

It is interesting to note that for the BMM choice, $f_i = f_i(O_i)$ [cf. Eq.(\ref{Obc1})],  and $\delta_i$ given by Eq.(\ref{eqAOS34}) together with the fact that on-shell we have
$O_b = m = O_c$, we find  that
\bqn
\lb{eq3.20ab}
   &&  \omega_{bb}^{\text{BMM}} =  -\frac{ \delta_b}{3 m}, \quad  \omega_{cc}^{\text{BMM}} =  -\frac{ \delta_c}{3 m},
   \quad \omega_{bc}^{\text{BMM}} = \omega_{cb}^{\text{BMM}} = 0, \nb\\
 &&
\Omega^{\text{BMM}}_b =  -\frac{ \delta_b}{3 m}, \quad \Omega^{\text{BMM}}_c =  -\frac{ \delta_c}{3 m}.
\eqn

To study the external spacetimes further,  in the following let us consider the two cases,    $\alpha_1 = 0$ and  $\alpha_1 \not= 0$, separately.

\subsection{External Spacetimes with $\alpha_1 \not=0$}

\begin{figure}[h!]
\includegraphics[width=0.9\linewidth]{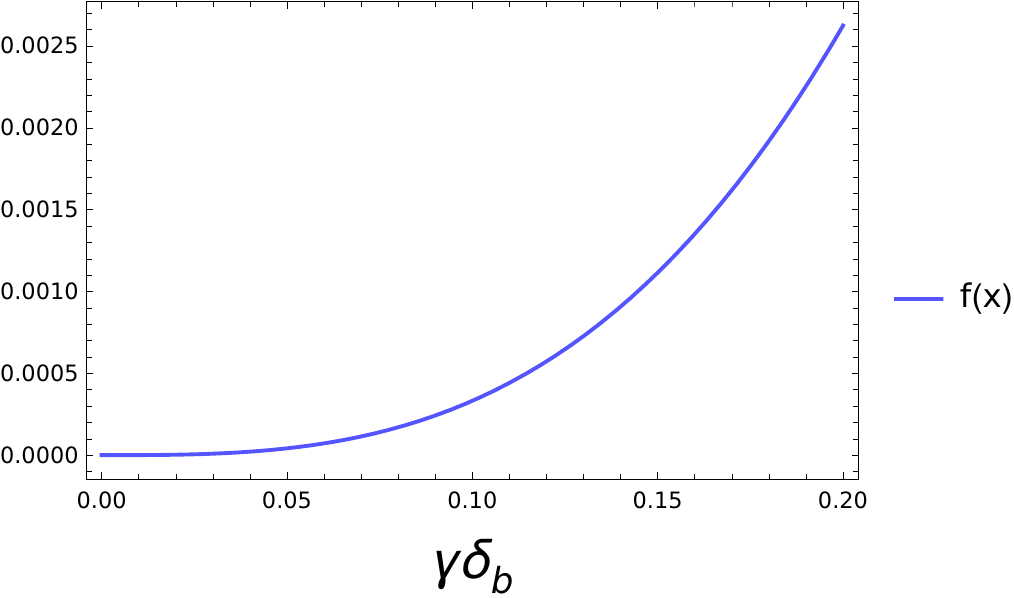}
\caption{The function $ f(x)$ defined by Eq.(\ref{fx})  vs
$x \equiv \gamma \delta_b$.} 
\label{f(x)}
\end{figure}

If $\alpha_1 \not= 0$, from Eq.(\ref{eq3.20}) we find that 
\begin{align}
\label{tbtcrel}
     t_c \approx \frac{\alpha_1}{b_o} e^{b_o t_b}.
\end{align}
Then, from Eq.(\ref{eq3.4}) we find that $dT = dt_b/C_{bc}$, and in terms of $t_b$ the metric (\ref{eq2.41})  becomes 
\begin{widetext}
\bqn
\label{temetric}
     ds^2 &=&-\frac{{p}_{b}^{2}}{\left|{p}_{c}\right| L_{o}^{2}} d x^{2}+\frac{\gamma^{2}\left|{p}_{c}\right| \delta_{{b}}^{2}}{\sinh ^{2}\left(\delta_{{b}} {b}\right) C_{bc}^2} d t_{b}^{2} 
      + \left|{p}_{c}\right|\left( d \theta^{2}+\sin ^{2} \theta d \phi^{2}\right),
\eqn
where $C_{bc}$ is given by Eq.(\ref{eq3.11}), and
\bqn
\lb{eq3.21}
    g_{xx} &\equiv& \frac{{p}_{b}^{2}}{\left|{p}_{c}\right| L_{o}^{2}} \simeq  \left(c_1 e^{2 b_o t_b} + c_2 e^{b_o t_b} + c_3 + \cdots \right)  \exp\left(- \frac{2\alpha_1}{b_o} e^{b_o t_b}\right),\nb\\
    g_{t_b t_b} &\equiv& \frac{\gamma^{2}\left|{p}_{c}\right| \delta_{{b}}^{2}}{\sinh ^{2}\left(\delta_{{b}} {b}\right) C_{bc}^2} \simeq  \left(d_1 e^{2 b_o t_b} + d_2 e^{b_o t_b} + d_3+ \cdots \right)   \exp\left(\frac{2\alpha_1}{b_o} e^{b_o t_b}\right),\nb\\
    g_{\theta\theta} &\equiv&  \left|{p}_{c}\right| \simeq 4 m^2 \exp\left( \frac{2\alpha_1}{b_o} e^{b_o t_b}\right), 
\eqn
where  ($c_i$, $d_i$) are constants defined as
\bqn
\lb{eq3.22}
c_{1} &\equiv& \frac{\left(b_o+1\right)^4}{16 b_o^4} , \quad  
    c_{2} \equiv -\frac{\left(b_o+1\right)^2}{4 b_o^4}, \quad
    c_{3} \equiv -\frac{\gamma ^2 \delta _b^2 \left(\gamma^2 \delta _b^2+4\right)}{8 b_o^4}, \nb\\
    d_{1} &\equiv& \frac{\omega _{bb}^2 f^2 m^4 }{\gamma^2  \delta _b^4 b_o^4} \left(b_o+1\right)^4 , \quad
   d_{2} \equiv \frac{4  \omega _{bb} f m^3}{\gamma^2 b_o^4  \delta _b^{4} } \left(b_o + 1\right)^{2}   \left\{\gamma b_o^2 \delta_b^2 
   - m\omega_{bb}\left(\gamma^3\delta_b^3 - b_o^2 f\right)\right\},\nb\\
   d_3 &\equiv& 2 m^2 \left(\frac{m \omega _{bb}}{\gamma ^2 \delta _b^4 b_o^4} \left(m \omega _{bb} \left(2 \gamma ^6 \delta _b^6+f^2 \left(\gamma ^4 \delta _b^4+4 \gamma ^2 \delta _b^2 \left(b_o^2+1\right)+8 b_o^2+2\right)+2 \gamma ^3 f \delta _b^3 \left(1-4 b_o^2\right)\right)\right.\right.\nb\\
   && ~~~~~~~~ \left.\left. +8 \gamma  f \delta _b^2 b_o^4 -4 \gamma ^4 \delta _b^5 b_o^2\right)+2\right),
\eqn
\end{widetext}
with
\bqn
\lb{fx}
f(\gamma\delta_b) &\equiv& b_o \cosh ^{-1}b_o -\gamma  \delta _b = b_o\ln\left(b_o + \gamma\delta_b\right) -\gamma  \delta _b \nb\\
&=& \frac{1}{3}\gamma^3\delta_b^3 + {\cal{O}}\left(\gamma^5\delta_b^5\right),
\eqn
which is always non-zero for $\gamma  \delta _b > 0$, as shown in Fig. \ref{f(x)}. 
 {The function $f(\gamma\delta_b)$ defined above must not be confused with the Dirac observables $f_i(O_b, O_c)\; (i = b, c)$  introduced in Eq.(\ref{eq3.1}).}

From the above expressions, it is clear that $\alpha_1$ must be positive,
in order to have the spacetime  {asymptotically} flat as $t_b \gg 1$. This is also consistent with Eq.(\ref{tbtcrel}), as we assumed that $t_b, \; t_c \gg 1$ asymptotically. Therefore, in the rest of this subsection we assume  $\alpha_1 > 0$, which requires
\bq
\lb{eq3.23a}
\Omega_{b}  > 0.
\eq
It is interesting to note that, corresponding to the BMM choices of $f_i = f_i(O_i)$ and $\delta_i$ given by Eq.(\ref{eqAOS34}), we have 
\bqn
\lb{eq3.20aba}
\Omega^{\text{BMM}}_b =  -\frac{ \delta_b}{3 m} < 0,  
\eqn
as given in Eq.(\ref{eq3.20ab}).  Therefore, the BMM choices cannot be realized in this case.  

It is also interesting to note that the spacetimes described by Eqs.(\ref{temetric})-(\ref{fx})  actually have similar asymptotic behavior as the AOS solution does, although the two metrics, given respectively by Eqs.(\ref{temetric}) and (\ref{eq2.41}),  look quite different.  
 To show this  claim, let us first introduce a new spacelike coordinate  $\xi$ via the relation, $\xi = e^{b_o t_b}$, and then we find that the metric (\ref{temetric}) becomes
 \begin{widetext}
\bqn
\label{eq3.23}
     ds^2 &\simeq&-\left(c_1 \xi^2 + c_2 \xi +  c_3 + \cdots\right) e^{-\alpha_0 \xi} dT^2 
      +  \left(d_1   + \frac{d_2}{\xi} + \frac{d_3}{\xi^2} + \cdots \right) e^{\alpha_0 \xi}\frac{d\xi^2}{b_o^2}
      + 4 m^2 e^{\alpha_0 \xi} \left( d \theta^{2}+\sin ^{2} \theta d \phi^{2}\right), ~~~~
\eqn
where $T \equiv x$ and $\alpha_0 \equiv {2\alpha_1}/{b_o} > 0$. 
Then, the corresponding curvature invariants of the above metric are given by
\bqn
\label{eq3.24}
     g^{\mu\nu}R_{\mu\nu} &\simeq& \left(\frac{2m}{r}\right)^2\left[\frac{1}{2}\left(\frac{1}{m^2}-\frac{b_o^2\alpha_0^2}{d_1}\right)+\frac{b_o^2\alpha_0(2d_1+d_2\alpha_0)}{2d_1^2\xi}+\mathcal{O}\left(\frac{1}{\xi^2}\right)\right], \nb\\
    R^{\mu\nu}R_{\mu\nu}&\simeq& \left(\frac{2m}{r}\right)^4\left[\frac{1}{8}\left(\frac{1}{m^4}+\frac{2b_o^4\alpha_0^4}{d_1^2}\right)-\frac{b_o^2\alpha_0(d_1^2+3b_o^2d_1m^2\alpha_0^2+b_o^2d_2m^2\alpha_0^3)}{2(d_1^3m^2)\xi}+\mathcal{O}\left(\frac{1}{\xi^2}\right)\right],\nb\\
      R^{\mu\nu\alpha\beta}R_{\mu\nu\alpha\beta}&\simeq&
      \left(\frac{2m}{r}\right)^4\left[\frac{d_1^2-2b_o^2d_1m^2\alpha_0^2+7b_o^4m^4\alpha_0^4}{4d_1^2m^4}+\frac{b_o^2\alpha_0^2(d_1d_2-b_o^2m^2\alpha_0(16d_1+7d_2\alpha_0))}{2d_1^3m^2\xi}+\mathcal{O}\left(\frac{1}{\xi^2}\right)\right],\nb\\
       C^{\mu\nu\alpha\beta}C_{\mu\nu\alpha\beta} &\simeq& \left(\frac{2m}{r}\right)^4\left[\frac{(d_1-4b_o^2m^2\alpha_0^2)^2}{12d_1^2m^4}+\frac{2b_o^2\alpha_0(2d_1+d_2\alpha_0)(d_1-4b_o^2m^2\alpha_0^2)}{3d_1^3m^2\xi}+\mathcal{O}\left(\frac{1}{\xi^2}\right)\right],
\eqn
\end{widetext}
where $r \left(\equiv 2m e^{\alpha_0 \xi/2}\right)$ is the geometric radius of the two spheres $\xi,\; T = $ constant. Comparing the above with the ones presented in \cite{AO20}, 
we can see that now the metric approaches asymptotically to the Minkowski spacetime as $r^{-4}$, which is the same as that of the AOS solution.

 It is also remarkable to note that for the AOS choices of $\delta_b$ and $\delta_c$ given by Eq.(\ref{eqAOS34}), we find that $\alpha_1 \propto m^{2/3}$ and $d_1 \propto m^{10/3}$.
  Then, the above expressions show that they are all independent of $m$ asymptotically.
   { In particular, we have 
 \bq
 \lb{asymptotic}
R^{\mu\nu\alpha\beta}R_{\mu\nu\alpha\beta} \simeq \frac{A_0}{r^4} + {\cal{O}}\left(\frac{1}{r^4 \xi}\right), 
 \eq
 where $\xi = \frac{2}{\alpha_0}\ln\left(\frac{r}{2m}\right)$, and $A_0$ is independent of $m$ given by
 \begin{equation}
\label{eqA0}
    A_0 \equiv \frac{28 \Omega _b^4}{\omega _{bb}^4}-\frac{8 \Omega _b^2}{\omega _{bb}^2}+4.
\end{equation}
This is sharply in contrast to the relativistic case, in which the Kretschmann scalar is given by 
 \bq
 \lb{GRK}
 \left. R^{\mu\nu\alpha\beta}R_{\mu\nu\alpha\beta}\right|_{\text{GR}} = \frac{48m^2}{r^{6}}.
 \eq
  It is also very interesting to note that the leading order of the Kretschmann scalar of the AOS solution  also behaves like $r^{-4}$ as $r \rightarrow \infty$ \cite{BBCCY20}. In the current case, even the Dirac observables $f_i$
  are chosen so that $A_0$ given by Eq.(\ref{eqA0}) is zero, the next leading order is ${\cal{O}}\left(\frac{1}{r^4 \xi}\right)$, which approaches zero still not as fast as  $r^{-6}$. In fact, it is even slower than 
  $r^{-5}$. }
  
   {To understand 
  the solutions further, we first note that to the leading order the metric takes the form
  \begin{equation}
    \label{eqmo}
    ds^2 \simeq - \frac{c_1 b_o^2}{\alpha_1^2} \left( \frac{\ln \left(\frac{r}{2m} \right)}{\frac{r}{2m}}\right)^2 dT^2 + \frac{d_1}{4m^2 \alpha_1^2} dr^2 + r^2 d\Omega^2.
\end{equation}
for $r \gg 2m$. On the other hand, the AOS solution takes the asymptotic form \cite{BBCCY20}
  \begin{equation}
    \label{eqmoB}
    ds_{\text{AOS}}^2 \simeq - r^{2(b_o-1)}dT^2 +   dr^2 + r^2 d\Omega^2,
\end{equation}
which is identical to the global monopole solution found in a completely different content \cite{BV89}.
 Then, the corresponding effective energy-momentum tensor  is given by
\bq
\lb{eqmoC}
T_{\nu\nu} = u_{\mu}u_{\nu} \rho + p_r r_{\mu}r_{\nu} + p_{\bot}\left(\theta_{\mu}\theta_{\nu} + \phi_{\mu}\phi_{\nu}\right),
\eq
where $u_{\mu}$ denotes the unit timelike vector along $T$-direction, and $r_{\mu}$,   $\theta_{\mu}$ and $\phi_{\mu}$ are the spacelike unity vectors along, respectively,
 $r-$, $\theta-$, and $\phi-$directions,   and
$\rho$, $p_r$ and $p_{\bot}$ are the energy density and pressures along the radial and tangential directions. To the leading order, they are given by
\bqn
\lb{eqmoD}
    \rho \simeq \frac{4 \alpha _1^2 m^2-d_1}{d_1 r^2}, \;\;
    p_r  \simeq -\frac{d_1+4 \alpha _1^2 m^2}{d_1 r^2}, \;\;
    p_{\bot}\simeq \frac{4 \alpha _1^2 m^2}{d_1 r^2},\nb\\
\eqn
which all approach zero as $r^{-2}$. This is also consistent with the asymptotical behaviors of the quantities  given in Eq.(\ref{eq3.24}). }

 {It should be also noted that, despite these differences, the spacetimes of the current solutions are also asymptotically
 flat and the corresponding ADM masses are as  well defined as that of the AOS solution \cite{AO20}. }

\subsection{External  Spacetimes with $\alpha_1 = 0$}

When $\alpha_1 = 0$,  from Eq.(\ref{eq3.20a}) and Fig. \ref{f(x)} we find that this can be the case  only  when
\begin{align}
\label{eq3.27}
    \Omega_{b} \equiv \omega_{bb} + \omega_{bc} = 0.
\end{align}
It is clear that the BMM choices of $f_i$ and $\delta_i$, given by Eqs.(\ref{Obc1}) and (\ref{eqAOS34}), are not compatible with this case, too.

Then,  to the leading order, Eq.(\ref{eq3.20}) yields
\begin{align}
    \label{eq3.28}
    t_b \simeq t_c \equiv t,
\end{align}
 as $t_c \rightarrow \infty$.  With Eqs.(\ref{eq3.27}) and (\ref{eq3.28}) we find that  
\bqn 
\label{eq3.28a}
    g_{xx}&\simeq&e^{-2t}\left(c_1e^{2b_ot}+c_2e^{b_ot}  + c_3 +\cdots\right)\nb\\
    g_{\theta \theta} &\simeq&  4 m^2 e^{2 t},
\eqn
where $c_n$'s are still given by Eq.(\ref{eq3.22}).
Finding the asymptotic limit of $g_{tt}$ is not so straightforward, and this is mainly because of the term $C_{bc}$ seen in the  expression
\begin{align}
     g_{tt} = \frac{\gamma^{2}\left|{p}_{c}\right| \delta_{{b}}^{2}}{\sinh ^{2}\left(\delta_{{b}} {b}\right) C_{bc}^2}.
\end{align}
The numerator of $C_{bc}$ in Eq.(\ref{eq3.12}) is equal to $1$ with the choice of $\omega_{bb}+\omega_{bc} = 0$ and the remainder of $g_{tt}$ is evaluated with the help of Mathematica, and is given by
\begin{align}
\begin{aligned}
g_{t t}  = e^{2 t}\Big( &d_{1} e^{2 b_o t}+d_{2} e^{b_o t} +d_3 + \cdots \Big),
  \end{aligned}
\end{align}
where  $d_n$'s are also given by Eq.(\ref{eq3.22}).
Introducing the new coordinates,  
\bq
\lb{eq3.28c}
    r=2m e^t, \quad x=\frac{4b_o^2}{(b_o+1)^2} \tau,
\eq
we find  
\begin{align}
\lb{eq3.28d}
    ds^2 = - g_{\tau\tau}d\tau^2 + g_{rr} dr^2 + r^2 d\Omega^2,
\end{align}
where
\bqn
\lb{eq3.28e}
g_{\tau\tau} &\simeq& \left(\frac{r}{2 m}\right)^{2\left(b_{o}-1\right)} \left(1 + \frac{c_2}{c_1} \left(\frac{2m}{r}\right)^{b_o}+ \frac{c_3}{c_1} \left(\frac{2m}{r}\right)^{2 b_o}\right), \nb\\
g_{r r}&\simeq& \frac{d_1}{4m^2} \left(\frac{r}{2m}\right)^{2 b_o} + \frac{d_2}{4m^2} \left(\frac{r}{2m}\right)^{b_o} + \frac{d_3}{4m^2}.
\eqn
Then, we find 
\begin{widetext}
\bqn
\label{eq3.29}
     g^{\mu\nu}R_{\mu\nu} &\simeq& \left(\frac{2m}{r}\right)^2\left[\frac{1}{2m^2}  
     +\frac{2(2b_o-1)}{d_1\xi^2} 
     +\mathcal{O}\left(\frac{1}{\xi^3}\right)\right], \nb\\
    R^{\mu\nu}R_{\mu\nu}&\simeq& \left(\frac{2m}{r}\right)^4\left[\frac{1}{8m^4}
    -\frac{b_o(c_1d_2-c_2d_1)}{2 c_1d_1^2m^2\xi^3}
    +\mathcal{O}\left(\frac{1}{\xi^4}\right)\right],\nb\\
      R^{\mu\nu\alpha\beta}R_{\mu\nu\alpha\beta}&\simeq&
      \left(\frac{2m}{r}\right)^4\left[\frac{1}{4m^4}  
      -\frac{2}{d_1m^2\xi^2} 
      +\mathcal{O}\left(\frac{1}{\xi^3}\right)\right],\nb\\
       C^{\mu\nu\alpha\beta}C_{\mu\nu\alpha\beta} &\simeq& \left(\frac{2m}{r}\right)^4\left[\frac{1}{12m^4}  
       +\frac{4(b_o-2)}{3d_1m^2\xi^2}+\mathcal{O}\left(\frac{1}{\xi^3}\right)\right],
\eqn
\end{widetext}
here $\xi \equiv (r/2 m)^{b_o}$.  {Interestingly the spacetime is again asymptotically flat}, and to the leading order has the same asymptotic behavior as that in the case $\alpha_1 \not= 0$.
 In particular, all these scalars are asymptotically independent of the mass parameter $m$, and approach zero as $r^{-4}$, 
  {sharply in contrast to the relativistic case given 
by Eq.(\ref{GRK}). However, different from the  case $\alpha_1 \not= 0$, to the next leader order the Kretschmann scalar behaves like ${\cal{O}}\left(1/r^{2(2+b_o)}\right)$.}

 {To study this case in more details, let us first note that to the leading order  the metric takes the form
\bqn
    \label{eqmoa10}
    ds^2 &\simeq& - c_1 \left(\frac{r}{2m} \right)^{2(b_o -1)}  dT^2 + \frac{d_1}{4m^2} \left(\frac{r}{2m} \right)^{2 b_o}  dr^2 \nb\\
    && ~~~~~~~ + r^2 d\Omega^2,
\eqn
 for $r \gg 2m$. Clearly, this is still different from Eq.(\ref{eqmoB}) for the AOS solution, despite the fact that to the leading order,
the Kretschmann scalar approaches zero like $r^{-4}$ in both cases. However, because the $r-$dependence of the $g_{rr}$ component, to the next leading order, 
 the Kretschmann scalar approaches zero like ${\cal{O}}\left(1/r^{2(2+b_o)}\right)$. Recall that $b_o \equiv \sqrt{1 + \gamma^2\delta_b^2} \ge 1$. This can be further understood by the analysis of the corresponding effective energy-momentum tensor,
 which can be also cast in the form of Eq.(\ref{eqmoC}), but now with    
\begin{align}
\lb{eqmoa10b}
    \rho &=-\frac{1}{r^2} \left({1+\frac{4 m^2 \left(2 b_o-1\right)}{d_1 \left(\frac{r}{2 m}\right)^{2 b_o}}}\right)  ,\nb\\
    p_r&=\frac{1}{r^2} \left({-1+\frac{4 m^2 \left(2 b_o-1\right)}{d_1 \left(\frac{r}{2 m}\right)^{2 b_o}}}\right) ,\nb\\
    p_{\bot}&=-\frac{4 m^2 \left(2 b_o-1\right)}{d_1 r^2 \left(\frac{r}{2 m}\right)^{2 b_o}},
\end{align}
which are consistent with the behaviors of the quantities given in Eq.(\ref{eq3.29}). Following \cite{AO20}, it is not difficult to see that  
the spacetimes of the current solutions are also asymptotically flat and the corresponding ADM masses are as  well defined as that of the AOS solution. }

\section{Canonical Phase Space Approach:   Internal Spacetimes} 
\label{Sec:IV}
 \renewcommand{\theequation}{4.\arabic{equation}}\setcounter{equation}{0}

In the internal region of the LQBH, the dynamical equations (\ref{eq3.2a}) and (\ref{eq3.2b}) take the form
\bqn
\lb{eq3.30a}
\frac{db}{dt_b} &=& - \frac{1}{2}\left(\frac{\sin\left(\delta_b b\right)}{\delta_b} + \frac{\gamma^2\delta_b}{\sin\left(\delta_b b\right)}\right), \\
\lb{eq3.30b}
\frac{dp_b}{dt_b} &=& \frac{1}{2}p_b\cos\left(\delta_b b\right)\left(1  -  \frac{\gamma^2\delta_b^2}{\sin^2\left(\delta_b b\right)}\right),
\eqn
for the variables $(b, p_b)$, and 
\bqn
\lb{eq3.31a}
\frac{dc}{dt_c} &=& - 2 \frac{\sin\left(\delta_c c\right)}{\delta_c}, \\
\lb{eq3.31b}
\frac{dp_c}{dt_c} &=&{2}p_c\cos\left(\delta_c c\right),
\eqn
for $(c, p_c)$. Eqs.(\ref{eq3.30a}) and (\ref{eq3.30b}) are identical with Eqs.(\ref{eq2.12}) and (\ref{eq2.13}), if we replace $T$ by $t_b$, while
Eqs.(\ref{eq3.31a}) and (\ref{eq3.31b}) are identical with Eqs.(\ref{eq2.14}) and (\ref{eq2.15}), if we replace $T$ by $t_c$. Then,
the corresponding solutions can be obtained directly from Eqs.(\ref{AOS2ba}) - (\ref{AOS2d}) by the above replacements, which lead to
\bqn
\lb{eq3.32a}
  \cos\left(\delta_b b\right) 
&=&  b_o\frac{1+ b_{o} \tanh\left(\frac{b_{o} t_b}{2}\right)}{b_o + \tanh\left(\frac{b_{o} t_b}{2}\right)} \nb\\
&=&   b_o \frac{b_+ e^{b_ot_b} - b_-}{b_+ e^{b_ot_b} + b_-}, \nb\\
p_b &=& - \frac{mL_o}{2b_o^2}\left(b_+  + b_- e^{-b_o t_b}\right){\mathfrak{A}},\\
\lb{eq3.32b}
 \sin\left(\delta_c c\right) &=& \frac{2a_oe^{2t_c}}{a^2_o + e^{4t_c}}, \nb\\
 p_c &=& 4m^2\left(a_o^2 + e^{4t_c}\right)e^{-2t_c}, 
\eqn
but now with  
\bqn
\lb{eq3.33}
{\mathfrak{A}} &\equiv& \Big[2\left(b_o^2 + 1\right)e^{b_o t_b} - b_-^2 - b_+^2 e^{2b_o t_b}\Big]^{1/2},
\eqn
where $a_o$ and $b_{\pm}$ are still given by Eq.(\ref{AOS2c}), and 
the range of the variables is given by Eq.(\ref{AOS2d}). 
 Then, it can be seen that the two Dirac observables $O_b$ and $O_c$ are also given by Eq.(\ref{eq2.25}) along the dynamical trajectories.
  However, instead of imposing the conditions (\ref{eqAOS34}), now we shall leave the choice of $\delta_b$ and $\delta_c$ open, as we did in the last section.
 Thus, the corresponding internal spacetimes are described by the metric
\bqn
\lb{eq3.34}
ds^2 &=& - N^2 dT^2 + \frac{p_b^2}{|p_c| L_o^2} dx^2 + |p_c|d\Omega^2\nb\\
&=& - \left(\frac{N}{C_{cb}}\right)^2 dt_c^2 + \frac{p_b^2}{|p_c| L_o^2} dx^2 + |p_c|d\Omega^2, ~~~
\eqn
where
\bqn
\lb{eq3.35}
N &\equiv& \frac{\gamma \delta_b\; \text{sgn}\left(p_c\right)\left|p_c\right|^{1/2}}{\sin\left(\delta_b b\right)}\nb\\
 &=& \frac{2m}{{\mathfrak{A}}}\left(b_+e^{b_o t_b} + b_-\right) \left(a_o^2e^{-2t_c} + e^{2t_c}\right)^{1/2}. ~~~~~~
\eqn
  
  In the following, let us study the above spacetimes near the horizons
  ($\mathfrak{A} = 0$) and throat ($\partial p_c/\partial t_c = 0$), separately. 
  
  \subsection{Spacetimes near the Horizons} 

 The horizons now are located at $\mathfrak{A} = 0$, which yields
 two solutions
 \bq\lb{eq3.36}
 t_b^{\text{BH}} = 0, \quad t_b^{\text{WH}} = - \frac{2}{b_o}\ln\left(\frac{b_o +1}{b_0 - 1}\right).
 \eq
 
 Now to find the relation between $t_b$ and $t_c$ the following expression has to be integrated 
 \bq
\lb{eq3.49}
dt_c = \frac{C_{cb}}{C_{bc}} dt_b,
\eq
 where the expressions of $C_{bc}$ and $C_{cb}$ in the interior are
  \bqn
  \lb{Cij}
 C_{b c}&=& \frac{1}{\cal{D}}\left(1- \Omega_{c}\frac{\partial O_{c}}{\partial \delta_{c}} \right),\nb\\
  C_{c b}&=& \frac{1}{\cal{D}}\left(1- \Omega_{b} \frac{\partial O_{b}}{\partial \delta_{b}} \right),
 \eqn
 but now with
\begin{widetext}
\bqn
\lb{eq3.51}
{\cal{D}} &\equiv& 1 - \omega_{cc} \frac{\partial O_{c}}{\partial \delta_{c}} -  \omega_{bb} \frac{\partial O_{b}}{\partial \delta_{b}} + \left( \omega_{bb}\omega_{cc}-\omega_{bc}\omega_{cb} \right) \frac{\partial O_{b}}{\partial \delta_{b}} \frac{\partial O_{c}}{\partial \delta_{c}}, \nb \\
\frac{\partial O_{b}}{\partial \delta_{b}} &=& -\frac{p_{b}}{2 \gamma L_{o}\delta_b^2}\left(1-\frac{\gamma^{2} \delta_{b}^{2}}{\sin ^{2}\left(\delta_{b}b\right)}\right) \big[\delta_{b} b \cos\left( \delta_{b} b\right)-\sin \left(\delta_{b}b\right)\big], \nb\\
\frac{\partial O_{c}}{\partial \delta_{c}}&=&\frac{p_{c}}{\gamma L_{o} \delta_c^2} \big[\delta_{c} c \cos\left( \delta_{c} c\right)-\sin\left( \delta_{c} c\right)\big].
\eqn
 \end{widetext}

Similar to the previous subsection,  {in the following section we consider the} cases $\alpha_1 = 0$ and $\alpha_1 \not=0$, separately. 

\subsubsection{$\alpha_1 =0$}
 \label{inta1=0}

 In this case, 
it is remarkable to note that by integrating Eq.(\ref{eq3.9}) we find the following explicit solution,
 \bqn
 \label{eq4.14}
 t_b &=& t_b^0 + t_c +  \frac{m\Omega_{c}}{\delta_c}
  \left\{\cosh\left(2{\cal{T}}\right)\tan^{-1}\left(e^{2{\cal{T}}}\right) \right.\nb\\
  && \left. - \cosh\left[2\left(t_c - {\cal{T}}\right)\right] \tan^{-1} \left[e^{-2\left(t_c - {\cal{T}}\right)}\right]\right\}, ~~~
 \eqn
 which holds for any $t_c$, including the region  $t_c \ge 0$, outside the black hole horizon, where $t_c = {\cal{T}}$ is the location of the transition surface, defined by Eq.(\ref{AOS2f}).  {And $t_b^0$ is an integration constant which will be set to zero in the following discussions}. When $t_c = 0$ the second term in the right-hand side of the above expression vanishes identically, 
 and as $t_c \rightarrow \infty$ it goes to zero as ${\cal{O}}\left(e^{-2t_c}\right)$. This is consistent with Eq.(\ref{eq3.20}).

\begin{figure}[h!]
\includegraphics[width=0.95\linewidth]{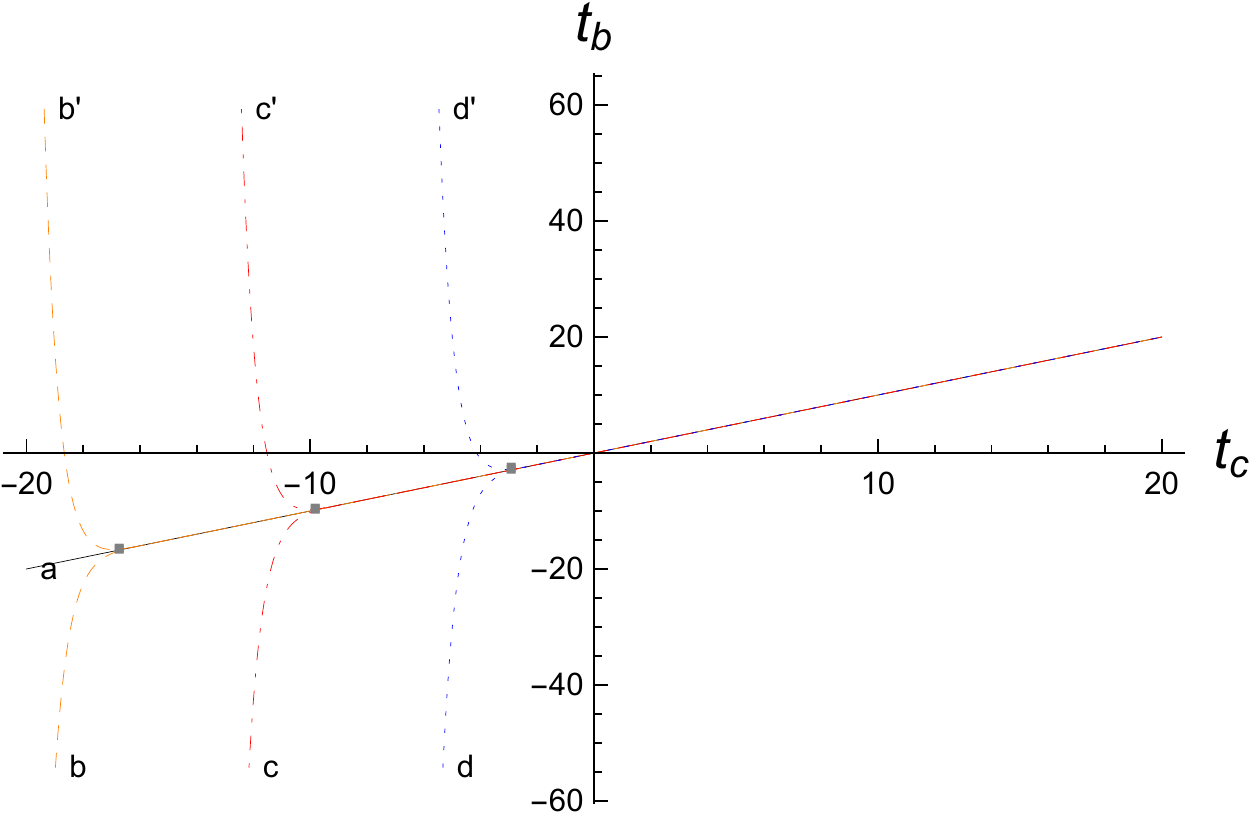}
\caption{Plots of  $t_b$ vs $t_c$   for $\alpha_1 = 0$ defined by Eq.(\ref{eq4.14}). Depending on the signs of $\Omega_c$, the dependence of $t_b$ on $t_c$ is different.  Curves $b$, $c$ and $d$  are all for  $\Omega_{c}=0.5$ but with different choices of $(m, \delta_c)$. In particular, they correspond to  $(m, \delta_c) = \{(10^6, 10^{-7}), (10^6, 0.1), (1, 0.1)\}$,  respectively. Curves  $b'$, $c'$ and $d'$ are all for $\Omega_c = -0.5$ {but with the same choice of $(m, \delta_c)$ as that of the unprimed curves in the respective order}.}
\label{tbvstc}
\end{figure}

 In Fig. \ref{tbvstc}, we plot  the curves of $t_b$ vs $t_c$ of Eq. (\ref{eq4.14}) for different choices of parameters involved. In particular, we find that the properties of $t_b$ across the transition surface sensitively depend on the signs of $\Omega_c$. More specifically, when $\Omega_c > 0$, $t_b$ decreases 
 exponentially right after crossing the transition surface, as $t_c$ becomes more and more negative, as shown by  Curves $b$, $c$ and $d$ with the choice $\Omega_c = 0.5$, where the  dots on the curves mark the locations of the transition surfaces. On the other hand, when $\Omega_c < 0$, $t_b$ 
 increases exponentially right after crossing the transition surface, as shown by  Curves $b'$, $c'$ and $d'$ with $\Omega_c =-0.5$. However, the locations of the transition surface indeed depend on the choices of the parameters $(m, \; L_0\delta_c)$, as shown by Eq.(\ref{AOS2f}). In particular, 
  Curves $b$, $c$ and $d$ {respectively} correspond to  
  $$
  \left(\frac{m}{m_p}, \frac{L_o \delta_c}{\ell_p}\right) =\left\{\left(10^6, 10^{-7}\right), \left(10^6, 0.1\right), \left(1, 0.1\right)\right\},
  $$
while Curves  $b'$, $c'$ and $d'$ are all for the same choices  of $(m, \delta_c)$, as {that of the unprimed curves in respective order}.
 Curves $b$ and $c$ share the same mass, i.e. $m/m_p = 10^6$, but with different $\delta_c$'s.
 Meanwhile, the locations of the  throats (the gray dots) move from the left-hand side to the right-hand side in the  direction closer to the horizon, which means that the quantum effects increase as $\delta_c$ increases. Curves $c$ and $d$ share the same $\delta_c=0.1$, but different masses. Comparing their throat positions, we find that the smaller mass also means the more significant quantum effects. On the other hand, outside the horizon, no matter what the parameters are,  
 $t_b\simeq t_c$, which is consistent with our previous conclusion for large $t_b$ and $t_c$, as shown by Eq. (\ref{eq3.28}). 
To understand this point further,  let us expand the above expression around the horizon, for which we find 
 \bqn
 \lb{eq4.15}
 t_b &=& \beta_1 t_c +    \beta_2 t_c^2
 + \beta_3 t_c^3 +  {\cal{O}}\left(t_c^4\right), 
 \eqn
where
\begin{align*}
\beta_1 &\equiv 1 +   \frac{m \Omega_{c}}{a_o \delta_c}\left[a_o + \left( a_o^2 -1 \right) \tan ^{-1}(a_o) \right],
\end{align*}
\begingroup
\allowdisplaybreaks
\begin{align}
\beta_2 &\equiv -   \frac{m \Omega_{c}}{a_o\left( a_o^2 +1 \right) \delta_c} \left[a_o \left( a_o^2 -1 \right) \right.\nb\\
 & \left. + \left( a_o^2 +1 \right)^2 \tan ^{-1}(a_o)\right], \nb\\
\beta_3 &\equiv   \frac{2m \Omega_{c}}{3a_o  \left( a_o^2 +1 \right)^2\delta_c}  \left[a_o + 6 a_o^3 + a_o^5  \right. \nb\\
& +  \left. \left( a_o^2 -1 \right) \left( a_o^2 +1 \right)^2 \tan ^{-1}(a_o) \right].
\end{align}
\endgroup

For macroscopic black holes, we have $m/m_p  \gtrsim M_{\odot} \simeq 10^{38}$, while the semi-classical limit requires $L_0 \delta_c \ll 1$. Then, expanding $\beta_n$ in terms
of $a_o$, we find that
\bqn
\lb{4.16}
  \beta_1 &=& 1 +   \frac{m \Omega_{c}}{a_o \delta_c}\left[a_o + \left( a_o^2 -1 \right) \tan ^{-1}(a_o) \right]  \nb \\
  &\simeq&  1 + \frac{\gamma^2L_o^2\delta_c \Omega_{c}}{48m} + {\cal{O}}\left(a_o^4\right) \simeq 1, \nb\\
   \beta_2 &=& -\frac{\gamma^2L_o^2\delta_c \Omega_{c}}{24 m}+ {\cal{O}}\left(a_o^4\right) \simeq 0,  \nb \\ 
    \beta_3 &=& \frac{\gamma^2L_o^2\delta_c \Omega_{c}}{18 m}+ {\cal{O}}\left(a_o^4\right) \simeq 0. 
 \eqn
Therefore, for macroscopic black holes, the relation $t_b\simeq t_c$ near the horizon is well justified. Then,  
we find that the metric components take the form
\begin{widetext}
\bqn
\lb{4.18}
g_{xx} &=& \frac{e^{2 t_c} \zeta \left(t_c\right)^2}{16 b_o^4 \left(a_o^2+e^{4 t_c}\right)}\left(\left(b_o-1\right) e^{-b_o t_c}+b_o+1\right)^2, \nb \\
g_{t_ct_c}=&&\!\!\!\!\!4 m^2 e^{-2 t_c}  \left(a_o^2+e^{4 t_c}\right)\frac{ \beta_-\left(t_c\right)^2}{\zeta \left(t_c\right)^2} \Bigg\{1-\omega _{cc} \frac{\partial O_c}{\partial \delta _c}+(\Omega _{c}\frac{\partial O_c}{\partial \delta _c}-1)m \omega _{bb} e^{-b_o t_c} \frac{\left(2 b_o^2 e^{b_o t_c}-\zeta \left(t_c\right)^2\right)}{2 \gamma  \delta _b^2 b_o^2 \zeta \left(t_c\right)}  \nb \\
&&\times\left[\gamma  \delta _b \zeta \left(t_c\right)-b_o \beta _-\left(t_c\right) \cos ^{-1}\left(\frac{b_o \beta _-\left(t_c\right)}{\beta _+\left(t_c\right)}\right)\right]\Bigg\}^2, \\
\lb{4.19}
\zeta (t_c) &\equiv&  \left[2 \left(b_o^2+1\right) e^{b_o t_c}-\left(b_o-1\right)^2-\left(b_o+1\right)^2 e^{2 b_o t_c}\right]^{1/2},  \nb \\
\beta_\pm(t_c) &\equiv& \left(b_o+1\right) e^{b_o t_c}\pm \left(b_o-1\right). 
\eqn
 \end{widetext}

\begin{figure}[h!]
\subfloat[]{%
  \includegraphics[width=0.9\linewidth]{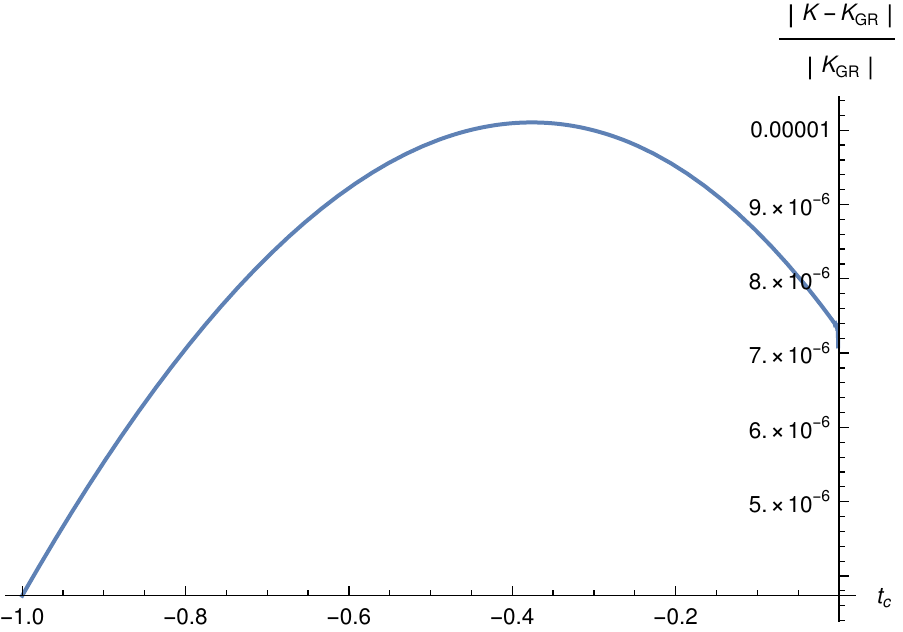}%
}\\
\subfloat[]{%
  \includegraphics[width=0.9\linewidth]{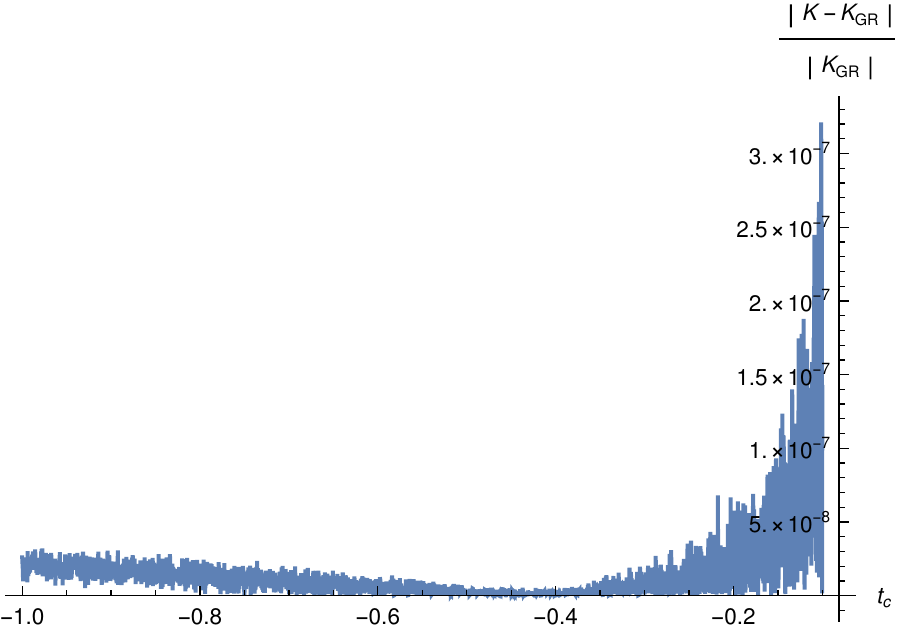}%
}
\caption{Plots of the relative difference between the Kretschmann scalars $K$ and $K_{\text{GR}}$ in the  $\alpha_1 =0$ case, for   (a)  $m =10^6$, and (b) $m =10^{12}$. 
Here $K_{\text{GR}} (\equiv 48m^2/p_c^3)$ is the corresponding  Kretschmann
scalar  given by GR.}
\label{fig4}
\end{figure}

To quantify the quantum effects near the horizon,  let us compute  the Hawking temperature at the horizon. Given a metric of the form
\bq
\lb{4.20}
ds^2 = - g_{tt}  dt^2 + g_{xx} dx^2 + p_c d\Omega^2,
\eq
the Hawking temperature of the black hole is given by \cite{AO20}, 
\bqn
\lb{4.21}
T_{H}=\frac{\hbar}{k_B \mathcal{P}}, \quad
\mathcal{P} = \lim_{t\to 0} \frac{4 \pi \left(g_{tt}g_{xx}\right)^{\frac{1}{2}}}{\partial_{t} g_{xx}},
\eqn
where $k_B$ is the Boltzmann constant.  Then, for the metric coefficients given by Eq.(\ref{4.18}) we find
\bqn
\lb{4.22}
T_{H}=  \frac{T_{H}^{\text{GR}}}{\left(1+a_o^2\right) \left(1+\varepsilon_T \right)},
\eqn
where $T_{H}^{\text{GR}} = \hbar/(8 k_B \pi m)$ denotes the Hawking temperature of the Schwarzschild black hole calculated in GR, and   
\bq
\lb{4.23}
\varepsilon_T \equiv \frac{m \omega _{cc}}{a_o \delta _c}\left[a_o+\left(a_o^2-1\right) \tan^{-1}\left(a_o\right)\right].
\eq
 For a BH of mass $10^6$, we find that 
 $$
 a_o^2 = \left(\frac{\gamma\delta_c L_o}{8m}\right)^2 \simeq 10^{-22},
 $$
  and
 $$
\varepsilon_T  =  \left(\frac{4 m \omega _{cc}}{3 \delta _c}\right)\left( 1- \frac{2}{5}a_o^2\right)a_o^2 + {\cal{O}}\left(a_o^6\right).
$$
For the AOS choice of Eq.(\ref{Obc1}), we find that  $4 m \omega _{cc}/(3 \delta _c) \simeq {\cal{O}}(1)$, so that
$\varepsilon_T  \lesssim 10^{-44}$,
that is, for macroscopic black holes, the quantum effects are negligible. This is consistent with what was concluded by AOS \cite{AOS18b,AO20}.

The above conclusion can be further verified by comparing the Kretschmann scalars $K$ with its relativistic counterpart  $K_{\text{GR}} \equiv 48m^2/p_c^3$.  {In particular, in Fig. \ref{fig4} we plot the relative difference of $K$ and $K_{\text{GR}}$
for $m = 10^6 m_{pl}$ and  $m = 10^{12} m_{pl}$, which indicate negligible quantum corrections near the horizon for massive LQBHs.}

\subsubsection{$\alpha_1 \not= 0$}

When $\alpha_1 \not= 0$, we find that
\begin{widetext}
\bqn
\lb{4.24}
 &&t_b \left(1+\frac{m \Omega _b}{\delta _b b_o^2}\right)+ \frac{2 m \Omega _b}{\delta _b b_o^2}-\frac{m \Omega _b}{2 \gamma  \delta _b^2 b_o^2}e^{-b_o t_b} \beta_+\left(t_b\right) \zeta\left(t_b\right) \cos ^{-1}\left(\frac{b_o \beta_-\left(t_b\right)}{\beta_+\left(t_b\right)}\right)\nb\\
&+&   \frac{m \Omega _b }{2 \delta _b b_o^3} \left(\left(b_o-1\right)^2 e^{-b_o t_b}-\left(b_o+1\right)^2 e^{b_o t_b}-2 b_o^3 t_b\right) + t_b^0 \nb \\
& =& t_c + \frac{m\Omega_{c}}{\delta_c}
  \left(\cosh\left(2{\cal{T}}\right)\tan^{-1}\left(e^{2{\cal{T}}}\right)  - \cosh\left(2\left(t_c - {\cal{T}}\right)\right) \tan^{-1} \left(e^{-2\left(t_c - {\cal{T}}\right)}\right)\right),
\eqn
  {where $t_b^0$ is an integration constant and will be set to zero as done previously in the $\alpha_1 = 0$ case}. 
\end{widetext}

\begin{figure}[h!]
\includegraphics[width=0.9\linewidth]{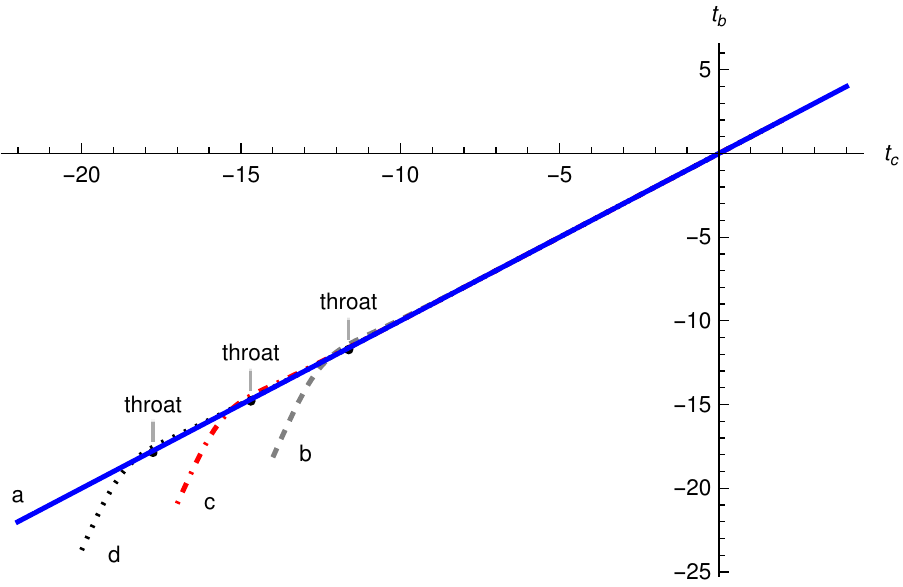}
\caption{Plots of Eq.(\ref{4.24}) for various choices as  given by Eq.(\ref{4.28}).}
\label{Fig4}
\end{figure}

\begin{figure}[h!]
\subfloat[]{%
  \includegraphics[width=0.9\linewidth]{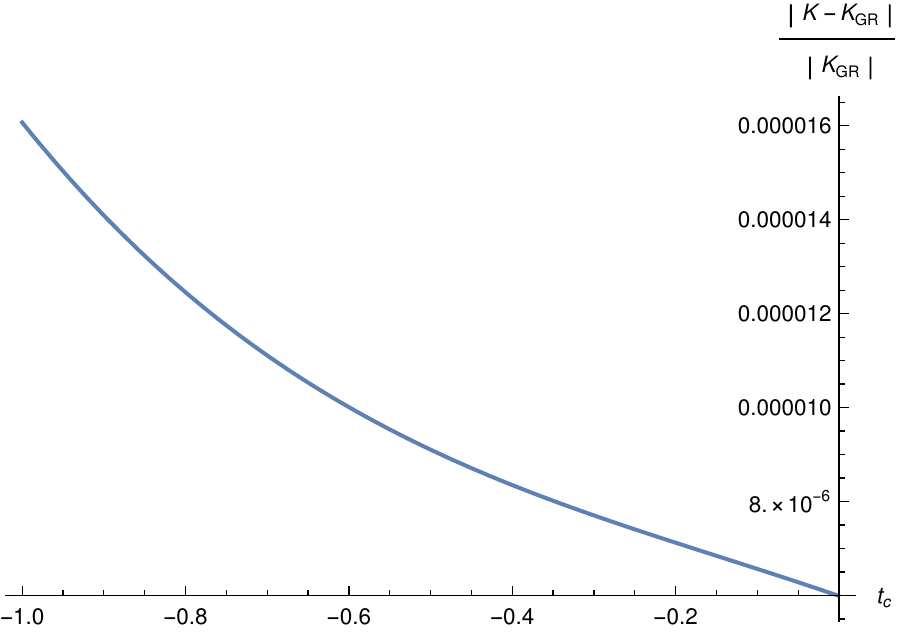}%
}\\
\subfloat[]{%
  \includegraphics[width=0.9\linewidth]{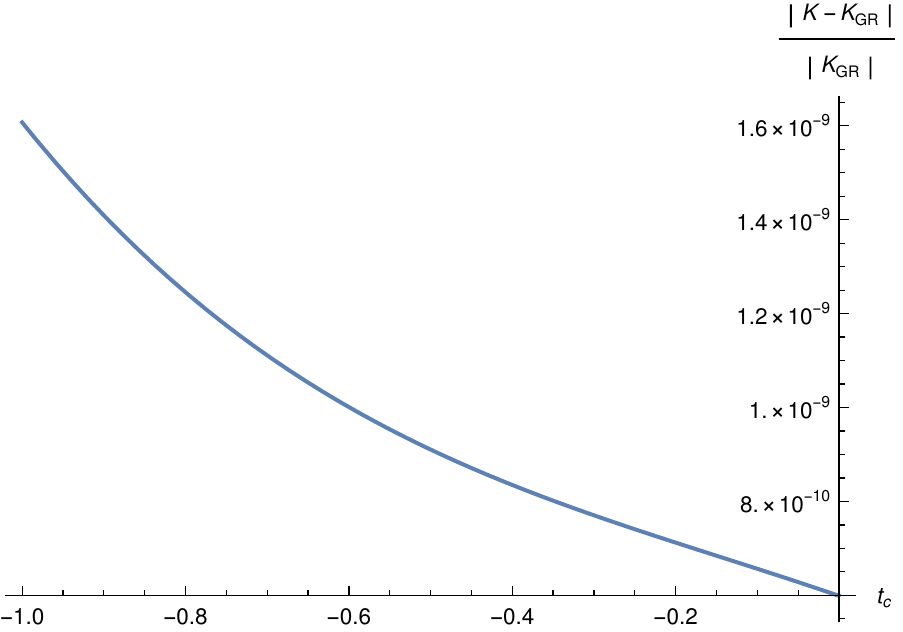}%
}
\caption{Plots of the relative difference between the  Kretschmann scalars $K$ and $K_{\text{GR}}$ in the  $\alpha_1 \neq 0$ case, for  (a)  $m =10^6$, and (b) $m =10^{12}$.}
\label{Fig6}
\end{figure}

Notice that the $t_c$ part of the above expression is precisely the right-hand side of Eq.(\ref{eq4.14}), and we showed explicitly in the last subsection that near the horizon $t_c =0$   the right-hand side can be well approximated by $t_c$. 
Now, expanding the $t_b$ part of the above expression around the horizon, we find
\bq
\lb{4.25}
t_b + \nu_2 t_b^2 + \nu_3 t_b^3 +  {\cal{O}}\left(t_b^4\right),
\eq
where 
\bqn
\lb{4.26}
\nu_2 &=& \frac{1}{6} m \gamma ^2  \delta _b \Omega _b, \nb \\
\nu_3 &=& \frac{1}{60}m \gamma ^2  \delta _b \Omega _b \left(10-\gamma ^2 \delta _b^2\right).
\eqn
The above coefficients $\nu_i$ are negligibly small for large black holes. For example, for a BH of mass $10^6$, they are of the order $\sim 10^{-9}$. Hence, for macroscopic black holes
Eq.(\ref{4.24}) can also be well approximated by
\bq
\lb{4.27}
t_b \simeq t_c,
\eq
near the black hole horizon,  similar to the case $\alpha_1 = 0$. This linear relation  can be confirmed by the plot of Eq.(\ref{4.24}) for various values, as seen in Fig. \ref{Fig4}. For plotting the curves b, c, and d corresponding to positive $\Omega_c$, the parameters are chosen  {respectively} as,   
\begin{align}
\lb{4.28}
    \left( \frac{m}{m_p} \right) &= \left(10^6, 10^8, 10^{10} \right) , \nb \\
     \left(\omega_{cc}, \omega_{cb}, \Omega_{c}\right) &= \left(\frac{\delta_c}{3 m}, 0, \frac{\delta_c}{3 m}\right),  \nb \\
    \left(\omega_{bb}, \omega_{bc}, \Omega_{b}\right) &= \left(\frac{\delta_b}{3 m},0, \frac{\delta_b}{3 m}\right),
\end{align}
where $\delta_i$'s are given by  Eq.(\ref{eqAOS34}).

Since in the current case ($\alpha_1 \not= 0$) $t_b \simeq t_c$ also holds  near the horizon for macroscopic black holes, the thermodynamics of the black hole horizon is quite similar to the case $\alpha_1 = 0$. In particular, its temperature is also given by Eqs.(\ref{4.22}) and (\ref{4.23}), and the difference to that of the Schwarzschild black hole calculated in GR is negligibly small for  macroscopic black holes.

 {Again, a plot of the relative difference between the Kretschmann scalar K and $K_{GR}$  is given in Fig. \ref{Fig6} for the $\alpha_1 \not= 0$ case, which also shows the negligible quantum effects near the horizons for massive LQBHs. }
 
\subsection{Spacetimes near  Transition Surfaces}

It is evident from Figs. \ref{tbvstc} and \ref{Fig4} that the above approximation, $t_b \simeq t_c$, is no longer valid once we start to probe the spacetime near and to the other side of the transition surface. We break this analysis again into two cases, $\alpha_1=0$ and $\alpha_1 \neq 0$.

\subsubsection{$\alpha_1 = 0$}

\begin{figure}[h!]
\centering
\includegraphics[width=0.9\linewidth]{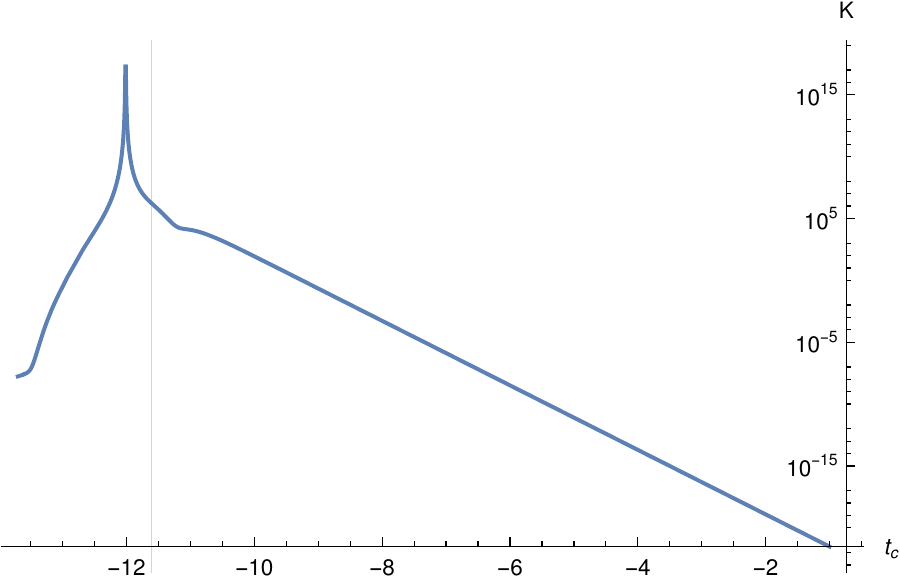}
\caption{The Kretschmann scalar near the transition surface $\tau_{ts} \simeq -11.6201$ denoted by the vertical line for the  case $\alpha_1=0$. Here   $m=10^6$, 
$(\omega_{cc}, \omega_{cb}) = (-{\delta_c}/{3 m}, 0)$, and $(\omega_{bb}, \omega_{bc}) 
=(-{\delta_b}/{3 m}, {\delta_b}/{3 m})$, so that $(\Omega_b, \Omega_c)  = (0,  -{\delta_c}/{3 m} < 0)$, where   $\delta_i$'s are given by Eq.(\ref{eqAOS34}).}
\label{Fig7}
\end{figure}

\begin{figure}[h!]
\centering
\includegraphics[width=0.9\linewidth]{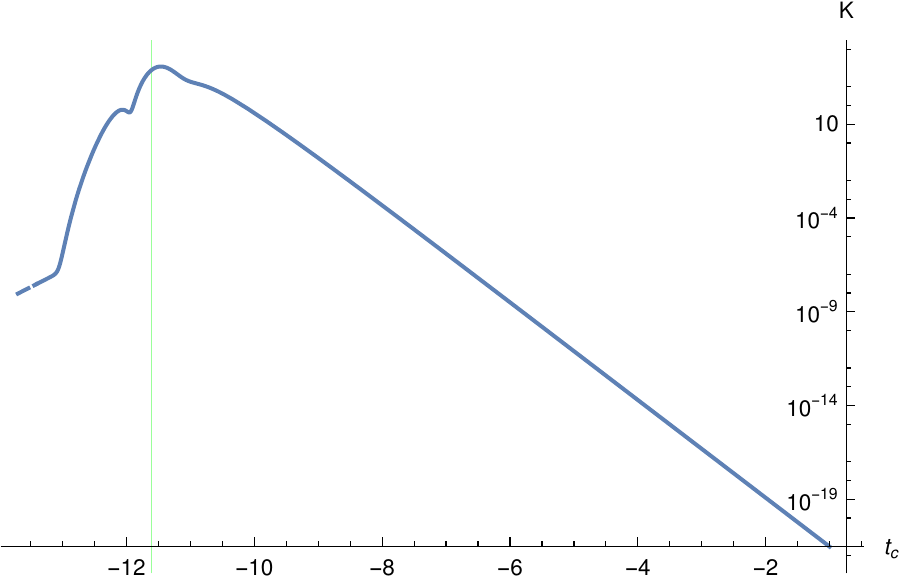}
\caption{The Kretschmann scalar near the transition surface  denoted by the vertical line for the  case $\alpha_1=0$. Here   $m=10^6$, $(\omega_{cc}, \omega_{cb}) = ({\delta_c}/{3 m}, 0)$, 
and $(\omega_{bb}, \omega_{bc}) 
=({\delta_b}/{3 m}, - {\delta_b}/{3 m})$, so that $(\Omega_b, \Omega_c)  = (0,  {\delta_c}/{3 m} > 0)$, where   $\delta_i$'s are given by Eq.(\ref{eqAOS34}).}
\label{Fig8}
\end{figure}

In this case, the relation between   $t_b$ and $t_c$ is given by Eq.(\ref{eq4.14}), which is valid everywhere in the interior. 
Combining this equation with the metric  (\ref{eq3.34}) we can calculate the curvature invariants to analyze the spacetime near the transition surface. We find that this can be done by  xAct\cite{xact}, a package for tensor computations in Mathematica, although the exact expressions are too complicated to be written down here. For this reason, we only plot out the Kretschmann scalar here for illustration,
 as other scalars like the Ricci scalar, Ricci tensor squared,  have similar features. In particular, in Figs. \ref{Fig7} and \ref{Fig8} we plot the Kretschmann scalar respectively for  $\Omega_c < 0$ and  $\Omega_c > 0$, but all with  $\Omega_b = 0$. In addition, we also provide Table \ref{Table1}, in which we show the explicit dependence of the maximal amplitude $K_m$ of the Kretschmann scalar on the mass $m$, the location of the  maximal amplitude of the Kretschmann scalar, denoted by $\tau_m$, and the location of the transition surface denoted by $\tau_{\text{ts}}$. To compare with the AOS solution, we also give the maximal amplitudes of the Kretschmann scalar for the AOS solution.

\begin{table}[h!]
\caption{The  maximal amplitude $K_m$ of the Kretschmann scalar $K$ for the  case $\alpha_1=0$ with different choices of the mass parameter $m$. Here $\tau_m$ denotes the location of the maximal point
of  $K$, and $\tau_{ts}$ the location of the  corresponding transition surface (throat).  To compare  it with that given by the AOS solution, we also give the maximal values of $K_m^{AOS}$.
Here we choose  $\omega_{bb} = -{\delta_b}/{3 m}$, $\omega_{bc} = {\delta_b}/{3 m}$, $\omega_{cc} =  -{\delta_c}/{3 m}$, and $\omega_{cb} = 0$, so that
$(\Omega_b, \Omega_c)  = (0,  -{\delta_c}/{3 m} < 0)$, where $\delta_i$'s are given by  Eq.(\ref{eqAOS34}).}
\begin{tabular}{|l|l|l||l|l|}
\hline
$m/m_p$  & $\tau_{m}$     & $K_m$    & $\tau_{ts}$  & $K_m^{AOS}$    \\ \hline
$10^6$    & -12.0147 & $2.46\times 10^{48}$ & -11.6201 & 82188.3628 \\
$10^8$    & -15.0848 & $1.56\times 10^{52}$ & -14.6902 & 82188.3642 \\
$10^{10}$ & -18.1549 & $3.60\times 10^{75}$ & -17.7603 & 82188.3642 \\
$10^{12}$ & -21.225  & $2.21\times 10^{75}$ & -20.8304 & 82188.3642 \\
$10^{14}$ & -24.2951 & $2.87\times 10^{70}$ & -23.9005 & 82188.3642 \\
$10^{16}$ & -27.3653 & $2.82\times 10^{69}$ & -26.9706 & 82188.3641 \\
$10^{18}$ & -30.4354 & $9.59\times 10^{77}$ & -30.0408 & 82188.3618 \\ \hline
\end{tabular}
\label{Table1}
\end{table}


  From Figs. \ref{Fig7} and \ref{Fig8} and Table \ref{Table1} we can see that the Kretschmann scalar remains finite across the transition surfaces, but  the maximal amplitude of the Kretschmann scalar sensitively depends on the mass $m$, which is in sharp contrast to the AOS solution in which  the maximal amplitude $K_m^{AOS}$ of the Kretschmann scalar remains the same \cite{AOS18a,AOS18b,AO20}.  
  
  Another unexpected feature is that {\it the maximal point of the  Kretschmann scalar 
 usually is not precisely at the transition surface, $\tau_m \not = \tau_{ts}$}.  {Although this looks strange,} a closer examination shows that this is due to two main facts: (1) the appearance of the factor $1/C_{bc}$ in the lapse function of the metric (\ref{eq3.34}),  and (2) the dependence of $t_b$ on $t_c$, which will lead to the modifications of $g_{xx}(t_b, t_c)$,  in comparison to the corresponding AOS component $g^{\text{AOS}}_{xx}(t_b, t_c)$ in which we have  $t_b = t_c = T$.

\begin{figure}[h!]
\centering
\includegraphics[width=0.9\linewidth]{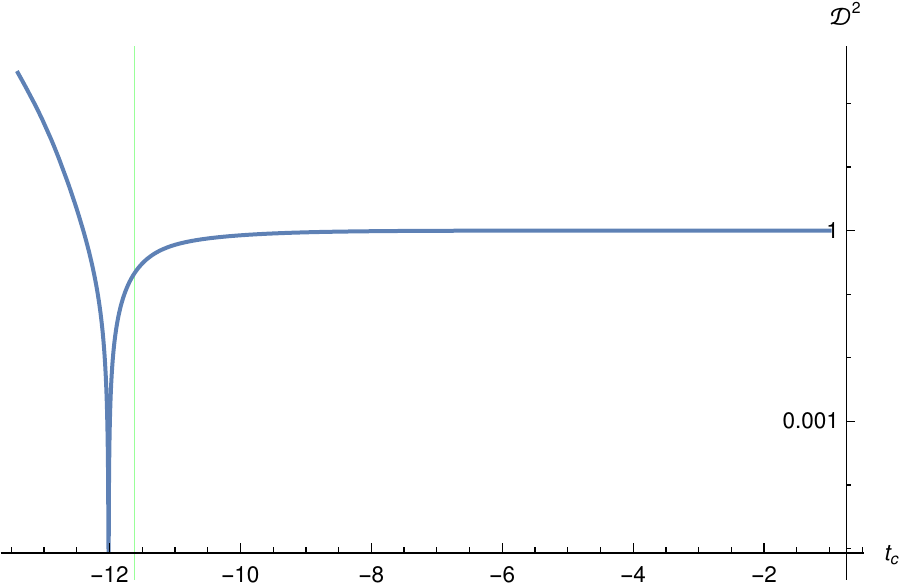}
\caption{The function ${\cal{D}}^2$ defined by Eq.(\ref{eq3.51}) for the case $\alpha_1 = 0$, for which we have $C_{bc} = 1/{\cal{D}}$. The vertical (green) line marks the position of the transition surface.
When plotting this curve, we have chosen the relevant parameters exactly as those given in Fig. \ref{Fig7}. }
\label{FigD}
\end{figure}

\begin{figure}[h!]
\centering
\includegraphics[width=0.9\linewidth]{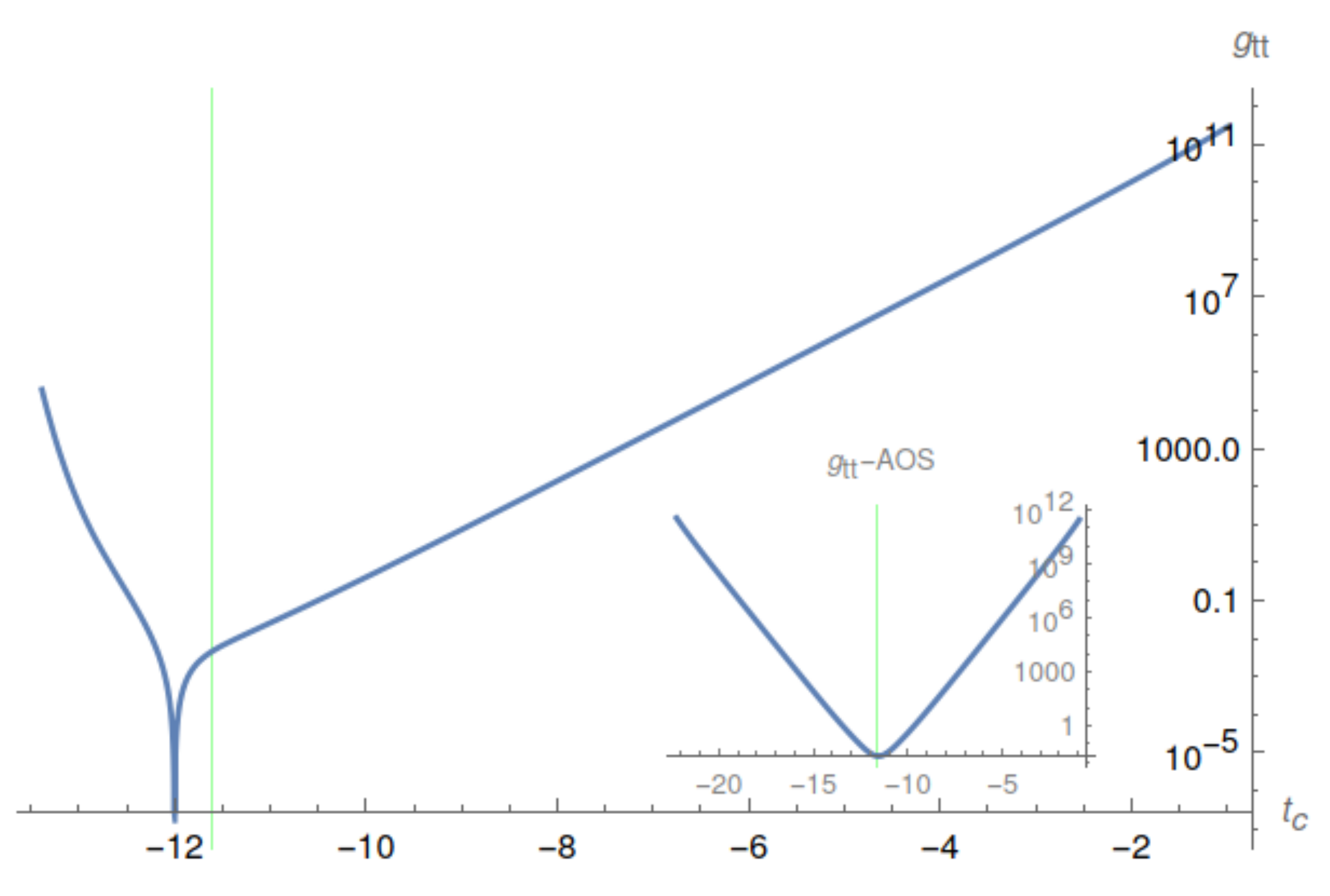}
\caption{The metric component $g_{t_ct_c}$ given in Eq.(\ref{eq3.34}), where $g_{tt} \equiv |g_{t_c t_c}|$. The inserting is the plot of the same quantity for the AOS solution. When plotting this curve, we have chosen the relevant parameters exactly as those given in Fig. \ref{Fig7}.}
\label{Fig12}
\end{figure}

\begin{figure}[h!]
\centering
\includegraphics[width=0.9\linewidth]{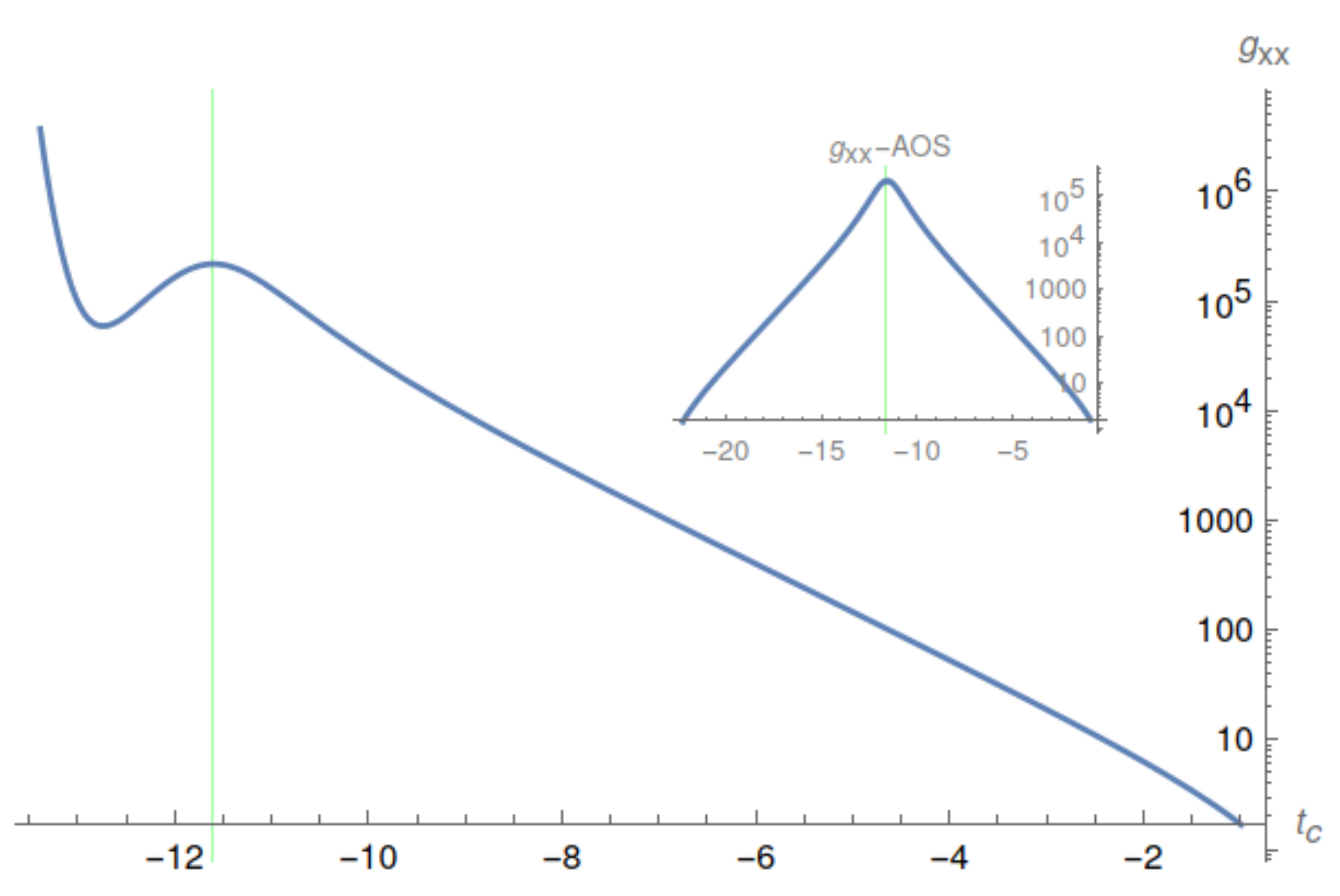}
\caption{The metric component $g_{xx}$ given in Eq.(\ref{eq3.34}). The inserting is the plot of the same quantity for the AOS solution. When plotting this curve, we have chosen the relevant parameters exactly as those given in Fig. \ref{Fig7}.}
\label{Fig13}
\end{figure}

 In particular, when $\alpha_1 = 0$, we have $1/C_{bc} = {\cal{D}}$, as can be seen from Eq.(\ref{Cij}), where ${\cal{D}}$ is defined by Eq.(\ref{eq3.51}). In Fig. \ref{FigD} we plot out the function ${\cal{D}}^2$
 for the same choices of the parameters as given in Fig. \ref{Fig7}, from which we can see that it changes dramatically near the maximal point $\tau_{m} \simeq -12.0147$ of the  Kretschmann scalar.
 In Figs. \ref{Fig12} and \ref{Fig13}, we plot out the metric components $g_{t_c t_c}$ and $g_{xx}$ given in Eq.(\ref{eq3.34}) vs $t_c$,  where $g_{tt} \equiv \left|g_{t_c t_c}\right|$. From these figures we can see
 clearly that both of these components change dramatically near the  maximal point $\tau_{m}$ of the  Kretschmann scalar. To compare it with the AOS solution, in each of these two figures, we also plot 
 the corresponding quantities for the AOS solution, from which it can be seen that no such behavior appears in the AOS solution.

  We also study the location of the white horizon and find that it is very near the transition surface. In particular, the ratio between WH and BH horizon radii now is much smaller than $1$ and sensitively depends on
 the mass parameter $m$, as shown explicitly in Table.\ref{Table2}. {Whereas in the AOS model this ratio is very close to $1$.}

\begin{table}[h!]
\centering
\caption{The ratio of the WH and BH horizon radii  for the  case $\alpha_1=0$ with different choices of the mass parameter $m$. Here we use the same choices as those in \ref{Fig8}, except for $m$.}
\begin{tabular}{|l|l|}
\hline
$m/m_p$  &    $\frac{r_{\scaleto{WH}{2pt}}}{r_{\scaleto{BH}{2pt}}}$ \\ \hline
$10^6$    & $5.5872 \times 10^{-5}$ \\
$10^8$    & $2.9462 \times 10^{-6}$ \\
$10^{10}$ & $1.5148 \times 10^{-7}$ \\
$10^{12}$ & $7.6577 \times 10^{-9}$ \\
$10^{14}$ & $3.8242 \times 10^{-10}$ \\
$10^{16}$ & $1.8923 \times 10^{-11}$ \\
$10^{18}$ & $9.2972 \times 10^{-13}$ \\ \hline
\end{tabular}
\label{Table2}
\end{table}

\subsubsection{$\alpha_1 \not= 0$}

\begin{figure}[h!]
\centering
\includegraphics[width=0.8\linewidth]{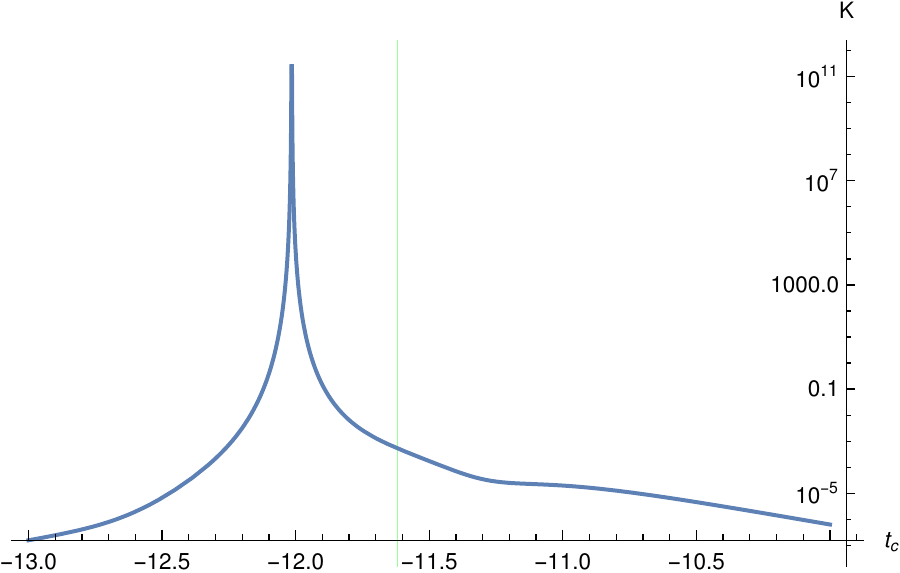}
\caption{The Kretschmann scalar for the case $\alpha_1 \not= 0$ with  $\Omega_c < 0$. In particular, the parameters are chosen as $m=10^6$, $(\omega_{cc}, \omega_{cb}, \Omega_{c}) = (-\frac{\delta_c}{3 m}, 0, -\frac{\delta_c}{3 m})$, $(\omega_{bb}, \omega_{bc}, \Omega_{b}) =(-\frac{\delta_b}{3 m},0, -\frac{\delta_b}{3 m})$, where $\delta_i$'s are given by  Eq.(\ref{eqAOS34}).}
\label{Fig10}
\end{figure}

\begin{figure}[h!]
\centering
\includegraphics[width=0.8\linewidth]{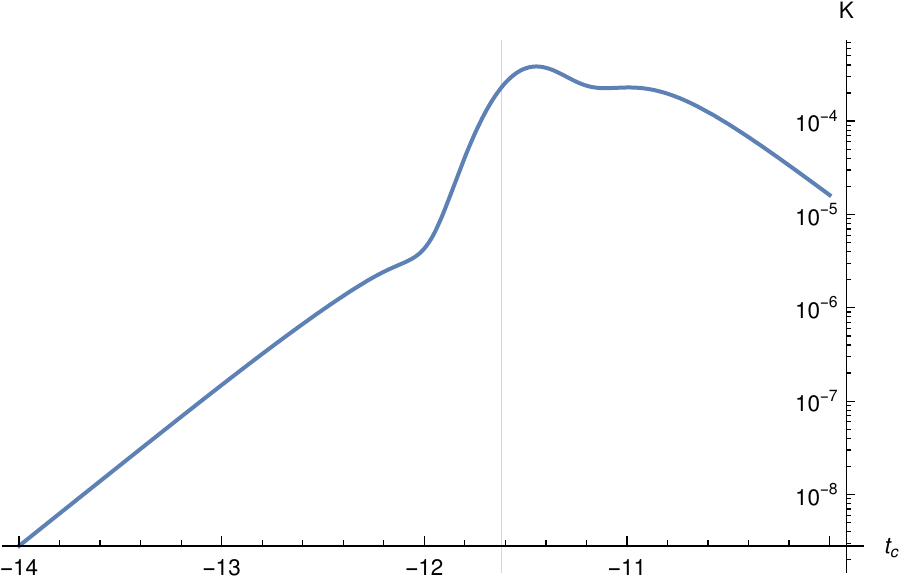}
\caption{The Kretschmann scalar for the case $\alpha_1 \not= 0$ with  $\Omega_c > 0$. In particular, the parameters are chosen as $m=10^6$, $ (\omega_{cc}, \omega_{cb}, \Omega_{c}) = (\frac{\delta_c}{3 m}, \frac{\delta_c}{3 m}, \frac{2\delta_c}{3 m})$, $(\omega_{bb}, \omega_{bc}, \Omega_{b}) =(\frac{\delta_b}{3 m},\frac{\delta_b}{3 m}, \frac{2\delta_b}{3 m})$, where $\delta_i$'s are given by  Eq.(\ref{eqAOS34}).}
\label{Fig11}
\end{figure}

In this case, the explicit relation between $t_b$ and $t_c$ is given by Eq.(\ref{4.24}). {This relation allows us to write down the metric and calculate the curvature invariants. Similar to the $\alpha_1 =0$ case, the exact expressions of them are too complicated to be written down explicitly here, and instead we find that it sufficient to simply plot them out.} Since they all have similar behavior, we plot out only the Kretschmann scalar. In particular, we plot it for $\Omega_c < 0$ and $\Omega_c > 0$, respectively in Figs. \ref{Fig10} and \ref{Fig11}. The vertical line in each of these figures represents the location of the transition surface, and is usually different from the maximal point of the  Kretschmann scalar, quite similar to the case $\alpha_1 = 0$ and for similar reasons. 

\section{Conclusions}
\label{Sec:V}
\renewcommand{\theequation}{5.\arabic{equation}}\setcounter{equation}{0}

In this paper, we  studied the 4-dimensional canonical  phase space approach, explored respectively by BMM \cite{BMM19} and GM \cite{GQ21} recently, in which the two  parameters 
$\delta_i\; (i = b, c)$ appearing in the  polymerization quantization \cite{Ashtekar20}
\bq
\lb{eq5.1}
b \rightarrow \frac{\sin(\delta_b b)}{\delta_b}, \quad c \rightarrow \frac{\sin(\delta_c c)}{\delta_c}, 
\eq
are considered as functions of the two Dirac variables $O_b$ and $O_c$ \cite{GQ21}
\bq
\lb{eq5.2}
\delta_i = f_i\left(O_b, O_c\right), \; (i, = b, c),
\eq
where  $O_b$ and $O_c$ are given by Eqs.(\ref{Ob}) and (\ref{Oc}). Note that BMM only considered the particular case $\delta_i = f_i\left(O_i\right)$ \cite{BMM19},  {the same as the AOS choice given in} Eq.(\ref{Obc2}), although AOS considered them in the extended 8-dimensional 
phase space $\Gamma_{\text{ext}}$. The corresponding dynamical equations are given by Eqs.(\ref{eq3.2aa}) and (\ref{eq3.2bb}), which allow analytical solutions in terms of $t_b$ and $t_c$, where $t_b$ and $t_c$ are
all functions of $T$ only, given by Eq.(\ref{eq3.4}).  

To  compare the AOS and BMM/GM approaches, in Section \hyperref[Sec:II]{II} we first presented the AOS model,  and  {discuss how to uniquely fix the} 
two Dirac observables $\delta_i$'s [cf. Eqs.(\ref{eqAOS3a}) and (\ref{eqAOS3b})] in  the extended phase space. In the large mass limit, these conditions lead to $\delta_i$'s given explicitly by Eq.(\ref{eqAOS34}). 

In the BMM/GM model, the black and white horizons, in general, all exist, and naturally divide the whole spacetime into the external and internal regions, where $T$ is timelike in the internal region and spacelike in the  external region. 
In Section \hyperref[Sec:III]{III},  {we briefly introduce} the BMM/GM approach and focused on studies of the external region of the spacetime. We found that the asymptotical flatness condition of the spacetime requires
\bq
\lb{eq5.3}
\Omega_{b} \ge 0,  
\eq
where $\Omega_{b}$ is defined in Eq.(\ref{eq3.13}), which excludes the BMM choice $\delta_i = f_i\left(O_i\right)$ \cite{BMM19}, for which we always have $\Omega^{\text{BMM}}_{b} < 0$, as shown explicitly by Eq.(\ref{eq3.20ab}).
Despite the significant difference of the metrics of the AOS and  BMM/GM models, we found that, to the leading order,  the asymptotical behavior of the spacetime in the two models  is universal and independent of the 
mass  parameter $m$ for the curvature invariants [cf. Eqs.(\ref{eq3.24}) and (\ref{eq3.29})].  {But, to the next leading order, they are different. In particular,  the Kretschmann scalar behaves as   
\bq
\lb{eq5.4}
 K \simeq \frac{A_0}{r^{4}} + {\cal{O}}\left(\frac{1}{r^{4}\xi}\right),
 \eq
  as $r \rightarrow \infty$, where $A_0$ is a constant and independent of $m$, and $r$ the geometric radius of the two-spheres. For the case $\alpha_1 \not= 0$, we have  $\xi = \frac{2}{\alpha_0}\ln\left(\frac{r}{2m}\right)$, and
  for $\alpha_1 = 0$, we have  $\xi = \left(\frac{r}{2m}\right)^{b_o}$. Here $\alpha_1$ is defined in Eq.(\ref{eq3.20a}). The differences from the next leading order can be understood more clearly from the 
  metric and the effective energy-momentum tensor, given, respectively,  by Eqs.(\ref{eqmo}), (\ref{eqmoD}), (\ref{eqmoa10}) and (\ref{eqmoa10b}). On the other hand, asymptotically the AOS solution takes the global monopole 
  form  (\ref{eqmoB}), found previously in a completely different content \cite{BV89}.}  Nevertheless, the leading behavior of the Kretschmann scalar in both cases  is  in sharp contrast to the classical case  \cite{BBCCY20,AO20}, 
  for which we have $K_{\text{GR}} = 48m^2/r^6$.

  In Section \hyperref[Sec:IV]{IV}, we conducted our studies on the internal region of the spacetime. We first showed that the quantum gravitational effects  {near} the black hole horizon are negligible for massive black holes, and both the  Kretschmann scalar and 
  Hawking temperature are indistinguishable from those of GR, as shown explicitly by Figs. \ref{fig4} and \ref{Fig6},  and Eq.(\ref{4.22}). However, despite the fact that  all the physical quantities are finite, and the Schwarzschild 
  black hole singularity is replaced by a transition surface whose radius is always finite and non-zero, the internal region near the transition surface is dramatically different from that of the AOS model in several respects:
  (1) First, the location of the maxima of the curvature invariants, such as the  Kretschmann scalar  {is displaced from the transition surface} as shown explicitly by Figs.\ref{Fig7}, \ref{Fig8},  \ref{Fig10} and \ref{Fig11}.  
  Detailed investigations of the metric components  reveal that this is  because 
  of the  {dependence of the two Dirac observables $\delta_i$'s on the 4D phase space of the Ashtekar variables ($b, c, p_b, p_c$)}, which considerably modifies   the structure of the spacetime. In particular, we plotted the metric components $g_{t_c t_c}$ and $g_{xx}$, respectively in Figs. \ref{Fig12} and \ref{Fig13},
  and then compared them with those given in the AOS model,  {where the maxima are always located} at the transition surface \cite{AOS18a,AOS18b,AO20}.  (2) The maxima of these curvature invariants depend on the choice of the
  mass parameter $m$. In particular, Table \ref{Table1} shows such dependence for the  Kretschmann scalar, which also shows that such dependence is absent in the AOS model. (3) {The location of the white hole horizon is very near to
   the transition surface, and  the ratio of the two horizon radii is much smaller than $1$, and depends sensitively  on $m$ as shown in Table.\ref{Table2}. All these results are significantly different from those obtained in  the AOS model.}

  In review of the results presented in this paper, it is clear that further investigations are highly demanded for LQBH models, in which the two polymerization  parameters $\delta_b$ and $\delta_c$ appearing in Eq.(\ref{eq5.1}) are considered as Dirac observables of the  4-dimensional phase space, spanned by ($b, p_b; c, p_c$), before accepting them as viable LQBH models  in LQG.
   In particular, in \cite{Giesel:2021rky} the consistent gauge-fixing conditions in polymerized gravitational systems were studied, and
  it would be very interesting to check how these conditions affect the results presented in this paper as well as results obtained in  other LQBH models. 
  

{\bf Notes-in-addition:} When we were finalizing our manuscript,   we came across three very interesting and  relevant articles \cite{NGM22a,GM22,NGM22b}.  {We will briefly comment on them here}. First, in \cite{NGM22a} the authors studied the physical 
meaning of the three integration constants, $C_1, C_2$ and $\bar{p}^0_c$, obtained from the integration of the three dynamical equations for the  variables $c, b$ and $p_c$, respectively, and found that $C_1$
 is related to the location of the transition surface, 
$C_2$ can be gauged away by the redefinition of time $t \rightarrow t + t_0$, where $t_0$ is a constant, while $\bar{p}^0_c$ is related to the mass parameter. 
A similar consideration was also carried out in \cite{GSSW2020} but for the BMM polymer black hole solution  \cite{BMM19a}. Second, in \cite{GM22} the authors studied the integrability of $G_b(t_b) \equiv 
\int^0_{t_b}F_{cb}(t'_b)dt'_b = \int^0_{t_c}F_{bc}(t'_c)dt'_c \equiv G_c(t_c)$,  the invertibility of $t_i = G_i^{-1}[G_j(t_j)]$, and the overlap of the images of $G_i$'s.  It was shown that $F_{ij}$'s are always integrable so that $G_i$'s always exist. 
The images of $G_i$'s can be always made overlapped by using the redefinitions of the two time-variables $t_i' = t_i + t_i^0$. In addition,  $t_i = G_i^{-1}$ is always invertible except at the zero points $G_i(t_i^{\text{ext}}) = 0$. Moreover, these
zero points never correspond to the same moment $T$, so at least one of the two $G_i$'s is invertible at any given moment $T$. It must be noted that all the studies carried out in \cite{GM22} were restricted to the internal region. 
When restricting our studies to this region, our results are consistent with theirs, whenever  the  problems of
 the integrability,   invertibility and overlap of the images, all studied in \cite{GM22}, are concerned.   Finally, in \cite{NGM22b}, the authors considered the quantization of the AOS extended phase space model, and found 
the conditions that guarantee the existence of physical states in the regime of large black hole masses, among other interesting results.

\section*{Acknowledgements}

We would like to thank G.A. Mena Marug\'an  and P. Singh for their valuable discussions, comments and suggestions. We also acknowledge the support provided by the Baylor University High Performance and Research Computing Services for carrying out the numerical computations of this paper. This work is supported in part by the Zhejiang Provincial Natural Science Foundation of China under Grant No. LR21A050001 and LY20A050002,  the National Key Research and Development Program of China Grant No.2020YFC2201503,  the National Natural Science Foundation of China under Grant No. 11675143 and No. 11975203, and the Fundamental Research Funds for the Provincial Universities of Zhejiang in China under Grant No. RF-A2019015.

\end{document}